\documentclass[aps,reprint,twocolumn,amsmath,amssymb,longbibliography]{revtex4-2}

\usepackage{bm}
\usepackage{graphicx}
\graphicspath{{./}}
\usepackage{hyperref}
\usepackage{booktabs}
\usepackage{xcolor}
\usepackage{mathtools}

\DeclareMathOperator{\Var}{Var}

\newcommand{\Ecent}{E_{\mathrm{cent}}}
\newcommand{\Edom}{E_{\mathrm{dom}}}
\newcommand{\RLH}{R_{\mathrm{LH}}}
\newcommand{\dEdc}{\Delta E_{\mathrm{dc}}}
\newcommand{\sigE}{\sigma_E}
\newcommand{\Iint}{I_{\mathrm{int}}}
\newcommand{\Veff}{V_{\mathrm{eff}}}
\newcommand{\Vs}{V_s}
\newcommand{\Vt}{V_t}
\newcommand{\Ws}{W_s}
\newcommand{\Wt}{W_t}
\newcommand{\xis}{\xi_s}
\newcommand{\nt}{n_t}
\newcommand{\at}{a_t}
\newcommand{\lopt}{\ell_{\mathrm{opt}}}
\newcommand{\Thetaeff}{\Theta_{\mathrm{eff}}}
\newcommand{\Da}{D_\alpha}

\begin{document}

\title{Peak-Decomposition-Free Inverse Metrology of \\
Hyperspectral Moir\'e Photoluminescence}

\author{Katsunori Wakabayashi}
\email{WAKABAYASHI.Katsunori@nims.go.jp}
\affiliation{Research Center for Materials Nanoarchitectonics (MANA),
National Institute for Materials Science (NIMS),
Namiki 1-1, Tsukuba 305-0044, Japan}

\date{\today}

\begin{abstract}
Hyperspectral photoluminescence (PL) of moir\'e transition-metal
dichalcogenide heterobilayers encodes spatially varying exciton landscapes,
but extracting that information is hampered by the ambiguity of multi-peak
spectral decomposition.  Here we develop a peak-decomposition-free inverse
framework for quantitative optical metrology of effective disorder
coordinates.  From the raw cube
$I(x,y,E)$ we construct physically motivated descriptor maps---centroid
energy, dominant emission energy, spectral width, low/high spectral-weight
ratio, and dominant--centroid offset.  Their spatial
autocorrelation hierarchy and covariance structure form a robust descriptor
fingerprint of multi-scale disorder-sensitive spectral statistics.  By matching
these descriptor summary statistics to
a minimal smooth-plus-trap generative model through a grid-Bayesian inverse,
we infer effective disorder coordinates $\Thetaeff=\{\Ws,\xis,\Wt,\nt\}$ with
explicit uncertainties.  Using synthetic PL cubes generated from controlled
multi-scale landscapes, we recover the well-identified smooth-disorder
coordinates and constrain the trap sector up to a strength--density degeneracy.
We report this degeneracy explicitly as an intrinsic identifiability limit
rather than a deficiency of the method, and map four canonical disorder regimes
onto a disorder-coordinate diagram.  The
descriptor statistics are stable against shot noise and pixel pitch, and behave
predictably under optical-resolution and energy-window changes.  The same
pipeline ingests experimental cubes without modification, making it directly
applicable to two-dimensional and moir\'e materials.  Our results establish
descriptor-based hyperspectral PL as a practical, minimal-assumption route to
optical disorder diagnostics and provide the validated core of a reusable
analysis workflow (\texttt{HyperPL-Diag}) for moir\'e exciton systems.
\end{abstract}

\maketitle

%%%%%%%%%%%%%%%%%%%%%%%%%%%%%%%%%%%%%%%%%%%%%%%%%%%%%%%%%%%%%
\section{Introduction}
\label{sec:intro}

Moir\'e superlattices in van der Waals heterobilayers~\cite{geim2013vdw,
novoselov2016heterostructures} host strongly correlated~\cite{wu2018hubbard,
regan2020mott,shimazaki2020strongly} and optically active
excitons~\cite{mak2010atomically,splendiani2010emerging,eda2011photoluminescence,%
chernikov2014exciton,berkelbachtheory,xiao2012coupled,mak2016photonics,%
schaibley2016valleytronics,regan2022emerging} whose properties vary on
micron and sub-micron scales~\cite{andrei2021marvels,mak2022semiconductor,
wang2018colloquium}.
In MoSe$_2$/WSe$_2$ and related heterostructures,
interlayer and moir\'e excitons~\cite{rivera2016valley,nagler2017interlayer,%
miller2017long,lindlau2018role,okada2018direct,alexeev2019resonantly,%
brem2020hybridized,bai2020excitons,shinokita2022valley,fang2023localization,%
blundo2024localisation,brotonsgisbert2024interlayer}
probe a landscape that is simultaneously shaped by long-wavelength strain and
dielectric inhomogeneity~\cite{conley2013bandgap,raja2019dielectric,
cadiz2017excitonic,moody2015intrinsic}, electrostatic fluctuations, the moir\'e
potential itself~\cite{yu2017moire,wu2017theory,naik2018ultraflatbands,%
tang2020simulation,shabani2021deep,kennes2021moire,yu2021moire}, and
short-range trap-like defects~\cite{komsa2015native,
parto2021defect,schaibley2022localized,seyler2019signatures,tran2019evidence,%
jin2019observation,rivera2015observation,ciarrocchi2022excitonic}.

Hyperspectral PL mapping,
which records a full emission spectrum $I(x,y,E)$ at every spatial pixel,
has emerged as a powerful probe of this spatial
inhomogeneity~\cite{ahmad2025hierarchical,alfrey2026revealing}.
The conventional route to interpreting such data is multi-peak fitting.
Each local spectrum is decomposed into a sum of Gaussian or Lorentzian
components that are then assigned to microscopic emission channels.  For
spatially varying moir\'e PL this approach is fragile.  The number of
spectral components is not known a priori, varies from pixel to pixel, and
the fitted parameters depend sensitively on initialization and on the
assumed lineshape.  The result is an analysis that is difficult to automate,
hard to reproduce, and cumbersome to compare across samples or processing
conditions.

A companion theoretical paper~\cite{wakabayashi2025descriptor} introduced a
descriptor-filter picture for hyperspectral moir\'e PL.  Its central idea is
that simple, peak-decomposition-free spectral descriptors---moments and
spectral-weight ratios computed directly from each spectrum---act as filters
that project different components of the multi-scale disorder landscape.
Different descriptors therefore exhibit different spatial correlation
lengths---a length scale long used to characterize disordered systems
broadly~\cite{anderson1958absence,abrahams1979scaling,lee1985disordered,%
kramer1993localization}---producing a
descriptor correlation hierarchy, and their mutual
covariances follow from the shape of the local spectra.  That work
\emph{explained} why a hierarchy appears.

\begin{figure*}[t]
  \centering
  \includegraphics[width=\textwidth]{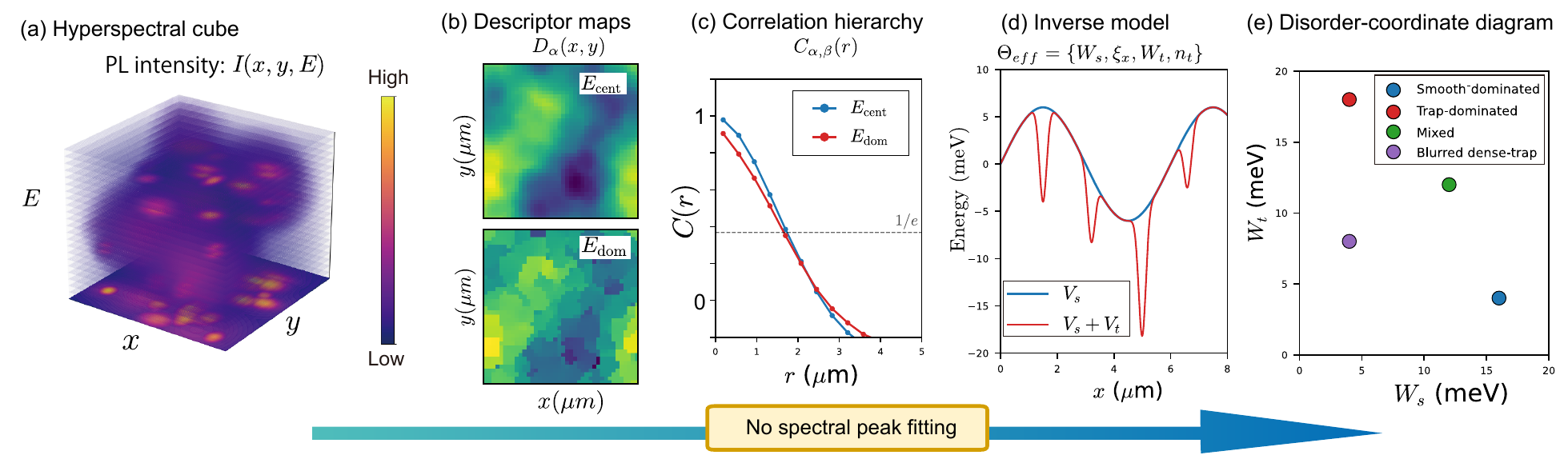}
  \caption{Concept and workflow of peak-decomposition-free inverse disorder
  metrology.  (a) The input is a hyperspectral PL cube $I(x,y,E)$, with two
  spatial axes $(x,y)$ and a photon-energy axis $E$.  The volume rendering shows
  the spatially varying, multi-peaked emission, color-coded by PL intensity,
  together with its projection onto the spatial plane.  The
  same pipeline accepts a synthetic cube (validation, this work) or an
  experimental cube (application).  (b) Peak-decomposition-free descriptor
  maps $\Da(x,y)$ are computed directly from each spectrum.  (c) Spatial
  autocorrelation and cross-correlation functions $C_{\alpha\beta}(r)$ yield
  a descriptor correlation hierarchy.  (d) A minimal smooth-plus-trap
  generative model with coordinates $\Thetaeff=\{\Ws,\xis,\Wt,\nt\}$ is the
  inverse engine.  (e) Samples or regions are placed on a disorder-coordinate
  diagram.  No step requires multi-peak decomposition.}
  \label{fig:workflow}
\end{figure*}

Here we reverse the logic.  Given a hyperspectral PL cube and its descriptor
statistics, what effective disorder coordinates are being optically observed?
We formulate this as an inverse problem and solve it without any multi-peak
decomposition.  Concretely, we (i) extract descriptor maps directly from the
cube, (ii) compute their spatial autocorrelation hierarchy and covariance,
(iii) compress these into a low-dimensional summary-statistics vector, and
(iv) match that vector to a minimal smooth-plus-trap generative disorder
model to infer effective disorder coordinates
$\Thetaeff=\{\Ws,\xis,\Wt,\nt\}$ with uncertainties.  The inverse is
transparent by design---physically interpretable summary statistics matched
to a generative model with an explicit posterior, rather than a black-box
regression---so that every inferred coordinate carries an auditable uncertainty
and its degeneracies are made explicit.  The full pipeline,
\begin{equation}
  I(x,y,E)\ \rightarrow\ \Da(x,y)\ \rightarrow\ C_{\alpha\beta}(r)\
  \rightarrow\ S_{\mathrm{desc}}\ \rightarrow\ \Thetaeff,
  \label{eq:pipeline}
\end{equation}
is summarized in Fig.~\ref{fig:workflow}, which runs from the input cube~(a)
through the descriptor maps~(b), their correlation hierarchy~(c), and the
generative inverse model~(d) to the resulting disorder-coordinate diagram~(e).

We adopt a synthetic-validation-first strategy.  Because the true disorder
parameters of a synthetic cube are known by construction, synthetic data let
us rigorously establish which descriptors respond to smooth versus trap
disorder, how optical averaging modifies apparent correlation lengths, which
parameters are robustly recoverable, and which are intrinsically degenerate.
This provides a stronger foundation than applying an unvalidated inverse
model directly to experimental data.  The framework is deliberately modular:
when raw experimental cubes become available, experimental descriptor maps
and summary statistics enter the same pipeline unchanged, and the synthetic
forward model serves as the inference engine and calibration layer.

The contribution of this paper is threefold.  First, we show that the
descriptor correlation hierarchy and covariance constitute a quantitative,
model-free descriptor fingerprint of disorder-sensitive spectral statistics.
Second, we demonstrate that a minimal
generative model converts this fingerprint into effective optical disorder
coordinates with explicit uncertainty quantification and identifiability
analysis---not a unique microscopic defect reconstruction.  Third, we establish that the
diagnostic descriptor statistics are stable against shot noise and pixel pitch and behave predictably
under optical-resolution and energy-window changes---the principal experimental
nuisances---making the framework suitable as the core of a reusable
diagnostic workflow.

%%%%%%%%%%%%%%%%%%%%%%%%%%%%%%%%%%%%%%%%%%%%%%%%%%%%%%%%%%%%%
%%%%%%%%%%%%%%%%%%%%%%%%%%%%%%%%%%%%%%%%%%%%%%%%%%%%%%%%%%%%%

\section{Data structure and descriptor extraction}
\label{sec:descriptors}

\subsection{Hyperspectral PL cube and preprocessing}

The input is a hyperspectral PL cube, that is, the continuous emission field
$I(x,y,E)$ of the Introduction sampled on a discrete grid $I(x_i,y_j,E_k)$, with
spatial pixels indexed by $i=1,\ldots,N_x$ and $j=1,\ldots,N_y$ and energy
channels by $k=1,\ldots,N_E$.  The descriptors below are written as energy
integrals for compactness and evaluated as sums over the measured channels.  Throughout, energy is
measured relative to a fixed reference $E_0$ taken at the exciton band edge,
so that $E<0$ denotes emission red-shifted below the band edge (e.g.\ from
trap-bound states).  In the synthetic validation we set $E_0=0$ without loss of
generality.  The absolute descriptors $\Ecent$ and $\Edom$ shift rigidly with
$E_0$, but the inference uses only shift-invariant descriptors ($\dEdc$, $\sigE$,
$\RLH$) and the spatial-fluctuation statistics of the maps (correlation
lengths, variances, and cross-correlations), all of which are invariant to the
choice of $E_0$.  In experimental applications $E_0$ may therefore be chosen as
a fixed global spectral reference, since the inference uses only these
shift-invariant descriptor combinations and spatial fluctuations.  

The minimal
metadata
required are the energy calibration, the spatial pixel pitch, an estimate of
the optical spot size or spatial resolution $\lopt$, and the excitation and
temperature conditions.  Preprocessing is intentionally simple and
transparent: dark/background subtraction, energy calibration, selection of an
energy window covering the relevant PL band, optional masking of dead or
extremely low-signal pixels, and---deliberately---no aggressive denoising in
the main analysis (denoising is treated as a robustness check in
Sec.~\ref{sec:robustness}).  Crucially, all descriptors below are computed
\emph{directly} from each spectrum, without fitting an assumed sum of peaks.

\subsection{Peak-decomposition-free descriptors}

For each spatial point $\bm{r}=(x,y)$ with local spectrum $I(\bm{r},E)$ we
define the primary descriptor set, summarized in
Table~\ref{tab:descriptors}.  The integrated intensity is
\begin{equation}
  \Iint(\bm{r}) = \int_{E_{\min}}^{E_{\max}} I(\bm{r},E)\,dE,
\end{equation}
sensitive to local radiative efficiency and exciton population.  The centroid
(first spectral moment) is
\begin{equation}
  \Ecent(\bm{r}) = \frac{\int E\,I(\bm{r},E)\,dE}{\int I(\bm{r},E)\,dE},
\end{equation}
expected to track smooth shifts from strain, dielectric, and electrostatic
disorder.  The dominant energy is
\begin{equation}
  \Edom(\bm{r}) = \arg\max_E I(\bm{r},E),
\end{equation}
sensitive to the strongest local emission channel and hence responsive to
traps and spectral fragmentation.  In experimental use $\Edom$ is stabilized
against shot noise and a discrete energy axis by mild spectral smoothing
together with a local three-point parabolic interpolation around the maximum.
In the synthetic analysis here we apply only the parabolic interpolation (no
smoothing).  Either way $\Edom$ remains peak-decomposition-free.  The spectral
width is the square root of
the second central moment,
\begin{equation}
  \sigE^2(\bm{r}) = \frac{\int [E-\Ecent(\bm{r})]^2 I(\bm{r},E)\,dE}
                         {\int I(\bm{r},E)\,dE}.
\end{equation}
With low- and high-energy windows split at $\Edom$, the low/high
spectral-weight ratio is
\begin{equation}
  \RLH(\bm{r}) = \frac{\int_{E<\Edom} I(\bm{r},E)\,dE}
                      {\int_{E>\Edom} I(\bm{r},E)\,dE},
  \label{eq:rhl}
\end{equation}
whose name matches its definition.  $\RLH$ grows as spectral weight shifts to
the low-energy side of the dominant peak (a low-energy trap tail).  Because
the split energy is the local $\Edom$, $\RLH$ measures the asymmetry around
the currently dominant emission channel rather than around a fixed global
energy.  In experimental applications a fixed reference (e.g.\ the global
spectral maximum or centroid) may be used instead, provided the convention is
kept fixed across samples.  The split choice does not affect the descriptor
maps or their individual spatial correlation lengths.  It does, however, set
the strength of the $\dEdc$--$\RLH$ co-variation reported below, which is
specific to the local-$\Edom$ split (Appendix~\ref{app:sensitivity}).
Finally, the dominant--centroid offset is
\begin{equation}
  \dEdc(\bm{r}) = \Edom(\bm{r}) - \Ecent(\bm{r}),
\end{equation}
a measure of spectral skewness.  All descriptors except $\Iint$ are
invariant to the overall intensity scale, so the analysis is insensitive to
absolute calibration of the detector.  We fix the sign and ratio conventions
explicitly here---$\RLH$ a low/high ratio and $\dEdc=\Edom-\Ecent$
(Table~\ref{tab:descriptors})---because they orient every cross-correlation
reported below.  The companion paper~\cite{wakabayashi2025descriptor} adopts
the opposite offset sign ($\Ecent-\Edom$) and a high/low ratio, so the
corresponding descriptor signs there are reversed.  Only magnitudes should be
compared across the two works.

\begin{table}[t]
\caption{Primary peak-decomposition-free descriptors and their expected
physical sensitivity.}
\label{tab:descriptors}
\begin{ruledtabular}
\begin{tabular}{lll}
Descriptor & Definition & Expected sensitivity \\
\colrule
$\Iint$   & $\int I\,dE$ & radiative efficiency \\
$\Ecent$  & $\int E I\,dE/\int I\,dE$ & smooth strain/dielectric \\
$\Edom$   & $\arg\max_E I$ & local traps \\
$\sigE$   & 2nd moment & fragmentation, broadening \\
$\RLH$    & low/high ratio & low/high weight balance \\
$\dEdc$   & $\Edom-\Ecent$ & skewness, low-E tail \\
\end{tabular}
\end{ruledtabular}
\end{table}

%%%%%%%%%%%%%%%%%%%%%%%%%%%%%%%%%%%%%%%%%%%%%%%%%%%%%%%%%%%%%
\section{Spatial descriptor statistics}
\label{sec:stats}

\subsection{Autocorrelation and correlation hierarchy}

For each descriptor $\Da(\bm{r})$ we form the fluctuation
$\delta\Da=\Da-\langle\Da\rangle$ and the normalized spatial autocorrelation
\begin{equation}
  C_{\alpha\alpha}(r) =
  \frac{\langle\delta\Da(\bm{R})\,\delta\Da(\bm{R}+\bm{r})
        \rangle_{\bm{R},|\bm{r}|=r}}
       {\langle[\delta\Da(\bm{R})]^2\rangle_{\bm{R}}}.
\end{equation}
Under spatial stationarity this autocorrelation is equivalent to the empirical
semivariogram
$\gamma(r)=\tfrac12\langle[\Da(\bm{R})-\Da(\bm{R}+\bm{r})]^2\rangle$ by
$\gamma(r)=\Var(\Da)\,[1-C_{\alpha\alpha}(r)]$.  We therefore estimate
$C_{\alpha\alpha}(r)=1-\gamma(r)/\Var(\Da)$ from the
variogram~\cite{cressie1993statistics} ($2\gamma$ is the structure function
familiar from random-media physics), a form that is numerically robust on
finite, masked maps because it requires no separate estimate of the spatial
mean.  The correlation length
$\xi_\alpha$ is obtained by a Gaussian fit
$C_{\alpha\alpha}(r)\simeq\exp[-r^2/(2\xi_\alpha^2)]$, with a $1/e$-crossing
fallback.  The hierarchy of correlation lengths,
\begin{equation}
  \{\xi(\Ecent),\,\xi(\Edom),\,\xi(\sigE),\,\xi(\RLH),\,\xi(\dEdc)\},
\end{equation}
is the central spatial observable.

\subsection{Cross-correlation and covariance}

For two descriptors the normalized cross-correlation $C_{\alpha\beta}(r)$ has
the zero-distance value $\rho_{\alpha\beta}=C_{\alpha\beta}(0)$, the Pearson
correlation coefficient between descriptor maps.  The most diagnostic pair is
$(\dEdc,\RLH)$.  For a unimodal spectrum dominated by a low-energy trap peak,
the centroid lies above the dominant energy, so $\dEdc=\Edom-\Ecent<0$.  Most
weight then sits on the high-energy side, so $\RLH$ (the low/high weight ratio)
is small, i.e.\ below unity.  As the
dominant peak shifts back toward the spectral center, both quantities increase
together.  The two descriptors therefore co-vary, giving a strongly
\emph{positive} $\rho(\dEdc,\RLH)$.  This provides a robust spectral-shape
correlation within the present descriptor convention---tied to the spectrum's
shape rather than to any particular sample---and its sign is fixed by that
convention.

%%%%%%%%%%%%%%%%%%%%%%%%%%%%%%%%%%%%%%%%%%%%%%%%%%%%%%%%%%%%%
\section{Minimal generative disorder model}
\label{sec:model}

\subsection{Effective potential}

We model the effective exciton landscape as
\begin{equation}
  \Veff(\bm{r}) = \Vs(\bm{r}) + \Vt(\bm{r}),
\end{equation}
a smooth field plus a trap field.  The smooth disorder is a Gaussian random
field with
\begin{equation}
  \langle\Vs(\bm{r})\Vs(\bm{0})\rangle = \Ws^2\,
  \exp\!\left(-\frac{r^2}{2\xis^2}\right),
\end{equation}
where $\Ws$ is the amplitude and $\xis$ the correlation length.  Physically it
represents the optically coarse-grained variation of the exciton energy from
strain, dielectric inhomogeneity~\cite{raja2019dielectric}, smooth
electrostatic fluctuations, and moir\'e/reconstruction-induced
modulation~\cite{weston2020atomic,mcgilly2020visualization}---not
the microscopic moir\'e potential, which lies below the optical resolution.  The trap
field is a Poisson set of attractive Gaussian wells,
\begin{equation}
  \Vt(\bm{r}) = \sum_i u_i\,
  \exp\!\left(-\frac{|\bm{r}-\bm{R}_i|^2}{2\at^2}\right),
\end{equation}
with positions $\bm{R}_i$ drawn from a homogeneous Poisson process of areal
density $\nt$, a common Gaussian width $\at$ (the spatial extent of each well),
and depths $u_i=-|\mathcal{N}(0,\Wt^2)|$, where
$\mathcal{N}(0,\Wt^2)$ denotes a zero-mean Gaussian (normal) variate of variance
$\Wt^2$, so $\Wt$ sets the characteristic trap depth.  These attractive wells
phenomenologically represent the localized, deeply trapping sites that give rise
to quantum-dot-like and single-photon emission in transition-metal
dichalcogenide (TMD) systems~\cite{koperski2015single,srivastava2015optically,he2015single,%
tonndorf2015single,branny2017deterministic}.
The trap radius $\at$
is fixed to a physically plausible value rather than treated as a free
parameter in the baseline inverse analysis.

\subsection{Optical resolution}

The measured signal is spatially averaged by the optical point-spread
function (PSF), modeled as a Gaussian whose $1$-sigma width relates to its full
width at half maximum (FWHM) by $\lopt=\mathrm{FWHM}/(2\sqrt{2\ln2})$.  Observed descriptor maps are therefore
optical-resolution-filtered maps, not microscopic disorder maps, and the
forward model includes this convolution explicitly so that inferred
correlation lengths are interpreted relative to $\lopt$.  Exciton
diffusion~\cite{kulig2018exciton,zipfel2020exciton}, if appreciable, provides an
additional spatial-averaging channel beyond the
optical PSF and can be absorbed phenomenologically into an effective resolution
length.  Separating the two requires independent calibration.

\subsection{Synthetic PL cube generation}

Each measurement pixel spectrum is generated as a broad smooth background plus
sharp trap peaks,
\begin{equation}
  I(\bm{R},E) = g_\Gamma\!\left[E-\bar{E}_s(\bm{R})\right]
  + A_t\!\!\sum_{i\in\mathrm{spot}}\! w_i(\bm{R})\,
  g_{\Gamma_t}\!\left[E-E_{t,i}\right],
  \label{eq:cube}
\end{equation}
where $g$ are normalized Gaussian lineshapes, $\bar{E}_s(\bm{R})$ is the
optical-spot average of $\Vs$ (the smooth background energy at pixel
$\bm{R}$), $E_{t,i}=\Vs(\bm{R}_i)+u_i$ is the ground energy of trap $i$, and
$w_i(\bm{R})$ is the optical-spot weight of trap $i$ at pixel $\bm{R}$.  This
minimal model deliberately reproduces descriptor statistics rather than every
microscopic peak.  The free coordinates are limited to
$\Thetaeff=\{\Ws,\xis,\Wt,\nt\}$, with $\at$, the broadenings
$\Gamma,\Gamma_t$, and $\lopt$ fixed or independently estimated.

%%%%%%%%%%%%%%%%%%%%%%%%%%%%%%%%%%%%%%%%%%%%%%%%%%%%%%%%%%%%%
\begin{figure*}[t]
  \centering
  \includegraphics[width=\textwidth]{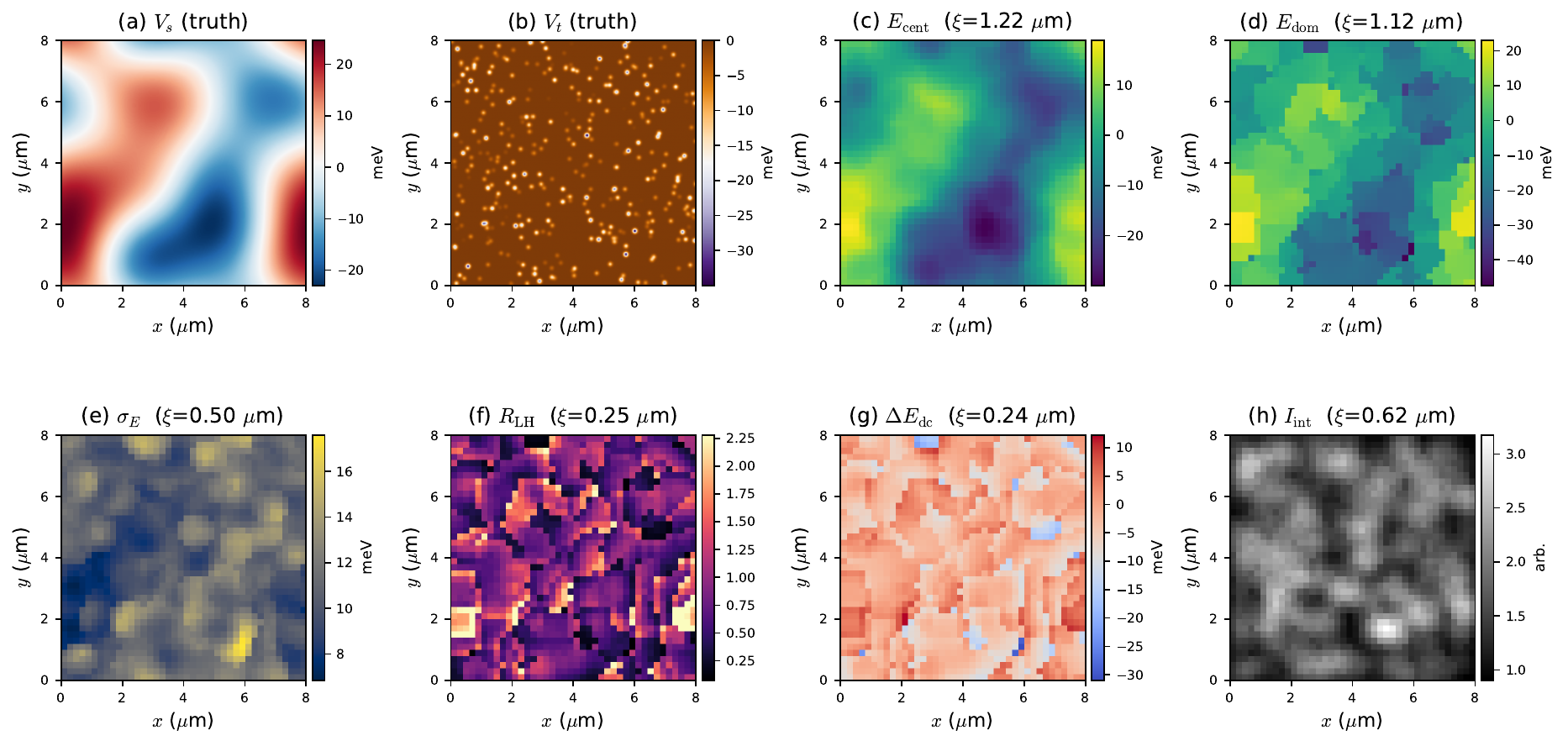}
  \caption{Descriptor maps from a single synthetic PL cube in the mixed
  regime ($\Ws=12$~meV, $\xis=2~\mu$m, $\Wt=12$~meV, $\nt=6~\mu$m$^{-2}$).
  (a,b) The true smooth and trap potentials.  (c--h) The
  peak-decomposition-free descriptor maps and their fitted correlation
  lengths.  The smooth-disorder-filtered descriptor $\Ecent$ (c) varies on
  long scales and tracks $\Vs$, whereas the trap-sensitive $\Edom$ (d),
  $\RLH$ (f), and $\dEdc$ (g) reveal short-scale structure; the spectral width
  $\sigE$ (e) and integrated intensity $\Iint$ (h) highlight trap-rich regions.
  Different descriptors expose different spatial structure from the same cube.
  $\Iint$ is shown for diagnostic completeness but is excluded from the baseline
  inverse (it is sensitive to excitation and collection efficiency).}
  \label{fig:maps}
\end{figure*}

%%%%%%%%%%%%%%%%%%%%%%%%%%%%%%%%%%%%%%%%%%%%%%%%%%%%%%%%%%%%%
\section{Synthetic disorder regimes and descriptor fingerprints}
\label{sec:regimes}

To map descriptor statistics onto disorder physics we study four
representative regimes that recur throughout the paper:
(i) a \emph{smooth-dominated} regime (large $\Ws$, long $\xis$, weak traps);
(ii) a \emph{trap-dominated} regime (strong traps on a weak smooth
background); (iii) a \emph{mixed} regime (comparable smooth and trap effects);
and (iv) a \emph{blurred dense-trap} regime (dense, shallow, sub-resolution
traps that mimic effective smoothing).  All cubes use experimental-scale
acquisition: an $8\times8\ \mu$m$^2$ area sampled on a $40\times40$ grid
(pitch $0.205\ \mu$m), an optical FWHM of $0.85\ \mu$m, and a $90$-meV energy
window.  Their ground-truth disorder coordinates are listed in
Table~\ref{tab:regimes}.

\begin{table}[t]
\setlength{\tabcolsep}{1.4pt}
\caption{Ground-truth disorder coordinates $\Thetaeff=\{\Ws,\xis,\Wt,\nt\}$ of
the four canonical synthetic regimes used throughout the paper.}
\label{tab:regimes}
\begin{ruledtabular}
\begin{tabular}{lcccc}
regime & $\Ws$ (meV) & $\xis$ ($\mu$m) & $\Wt$ (meV) &
  $\nt$ ($\mu$m$^{-2}$) \\
\colrule
smooth-dominated   & 16  & 2.5 & 4  & 2 \\
trap-dominated     & 4   & 1.0 & 18 & 6 \\
mixed              & 12  & 2.0 & 12 & 6 \\
blurred dense-trap & 4   & 1.0 & 8  & 10 \\
\end{tabular}
\end{ruledtabular}
\end{table}

Figure~\ref{fig:maps} shows the descriptor maps for the mixed regime, computed
from the same cube whose true smooth and trap potentials appear in panels~(a)
and~(b).  The centroid $\Ecent$~(c) varies smoothly and follows $\Vs$, while
$\Edom$~(d), $\RLH$~(f), and $\dEdc$~(g) display short-scale, trap-correlated
structure, and the spectral width $\sigE$~(e) and integrated intensity
$\Iint$~(h) highlight trap-rich regions.  This is a direct, qualitative
confirmation that distinct descriptors filter distinct components of the same
landscape.

\begin{figure*}[t]
  \centering
  \includegraphics[width=\textwidth]{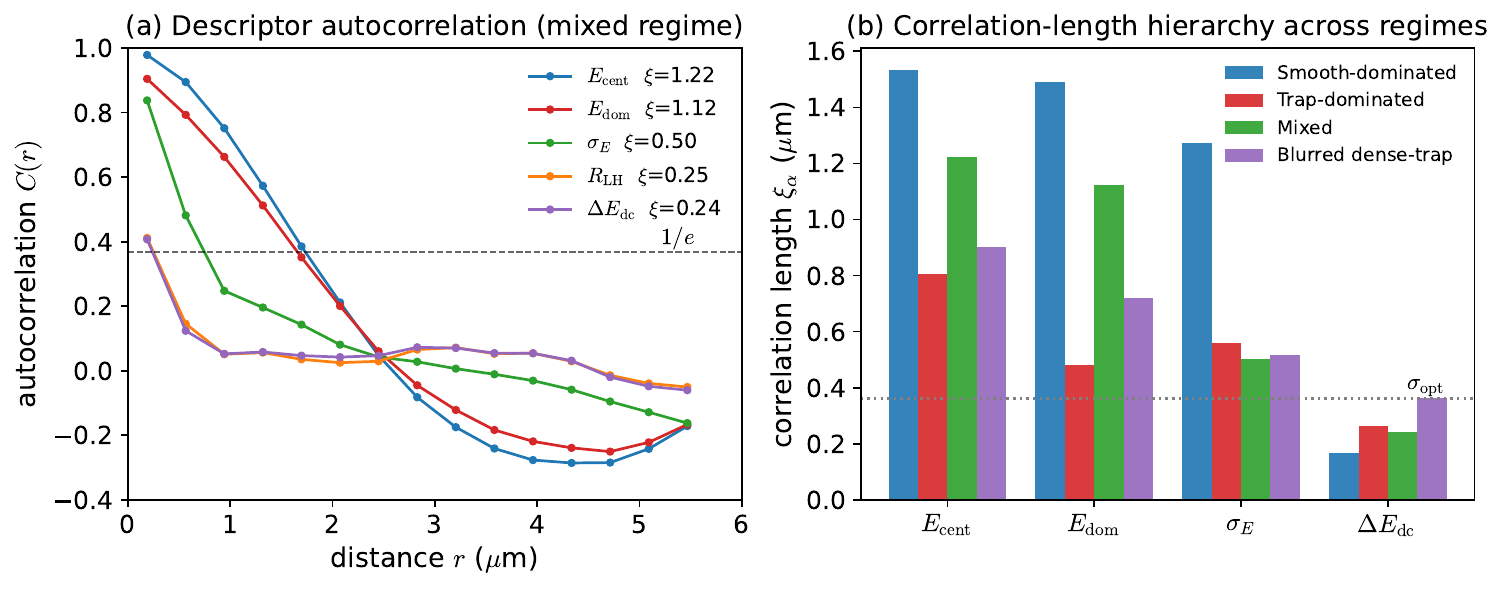}
  \caption{Descriptor correlation hierarchy.  (a) Autocorrelation functions
  $C(r)$ for the mixed regime with fitted correlation lengths; the dashed line
  is the $1/e$ level.  (b) Correlation-length hierarchy across the four
  regimes.  The ordering $\xi(\Ecent)\geq\xi(\Edom)$ is robust within the
  present smooth-plus-trap model across the regimes studied here, while its
  magnitude is regime-specific; the dotted line marks the optical-resolution
  scale $\sigma_{\rm opt}$, below which correlation lengths are interpreted
  cautiously.}
  \label{fig:hierarchy}
\end{figure*}

Figure~\ref{fig:hierarchy}(a) shows the descriptor autocorrelation functions
for the mixed regime.  The smooth-filtered $\Ecent$ has the longest correlation
length and the skewness descriptor $\dEdc$ the shortest.  Across all four
regimes [Fig.~\ref{fig:hierarchy}(b)] the ordering
$\xi(\Ecent)\!\geq\!\xi(\Edom)$ holds, but its magnitude is strongly
regime-dependent.  The ratio $\xi(\Ecent)/\xi(\Edom)$ is near unity in the
smooth-dominated regime and largest in the trap-dominated regime, where
$\Edom$ switches abruptly between traps while $\Ecent$ remains tied to the
smooth background.  The pattern of correlation lengths---not any single
value---is the diagnostic fingerprint.

This ordering and its regime
dependence persist across five correlation-length definitions and under
low-signal and random pixel rejection (Appendix~\ref{app:sensitivity}).  We
also recover the robust
spectral-shape correlation $\rho(\dEdc,\RLH)\approx+0.88$ across regimes;
its magnitude is comparable to the companion
theory~\cite{wakabayashi2025descriptor} and to the experimental
$|\rho|\approx0.98$~\cite{ahmad2025hierarchical}\footnote{Only the magnitude
should be compared across works.  The sign of $\rho(\dEdc,\RLH)$ is fixed by
the descriptor conventions of Sec.~\ref{sec:descriptors}
($\RLH$ a low/high ratio, $\dEdc=\Edom-\Ecent$), which may differ from those
used in the experimental and companion analyses.  Appendix~\ref{app:sensitivity}
quantifies how the $\RLH$ split convention sets this correlation.}.

%%%%%%%%%%%%%%%%%%%%%%%%%%%%%%%%%%%%%%%%%%%%%%%%%%%%%%%%%%%%%

%%%%%%%%%%%%%%%%%%%%%%%%%%%%%%%%%%%%%%%%%%%%%%%%%%%%%%%%%%%%%
\section{Inference of effective disorder coordinates}
\label{sec:inference}

\begin{figure*}[t]
  \centering
  \includegraphics[width=\textwidth]{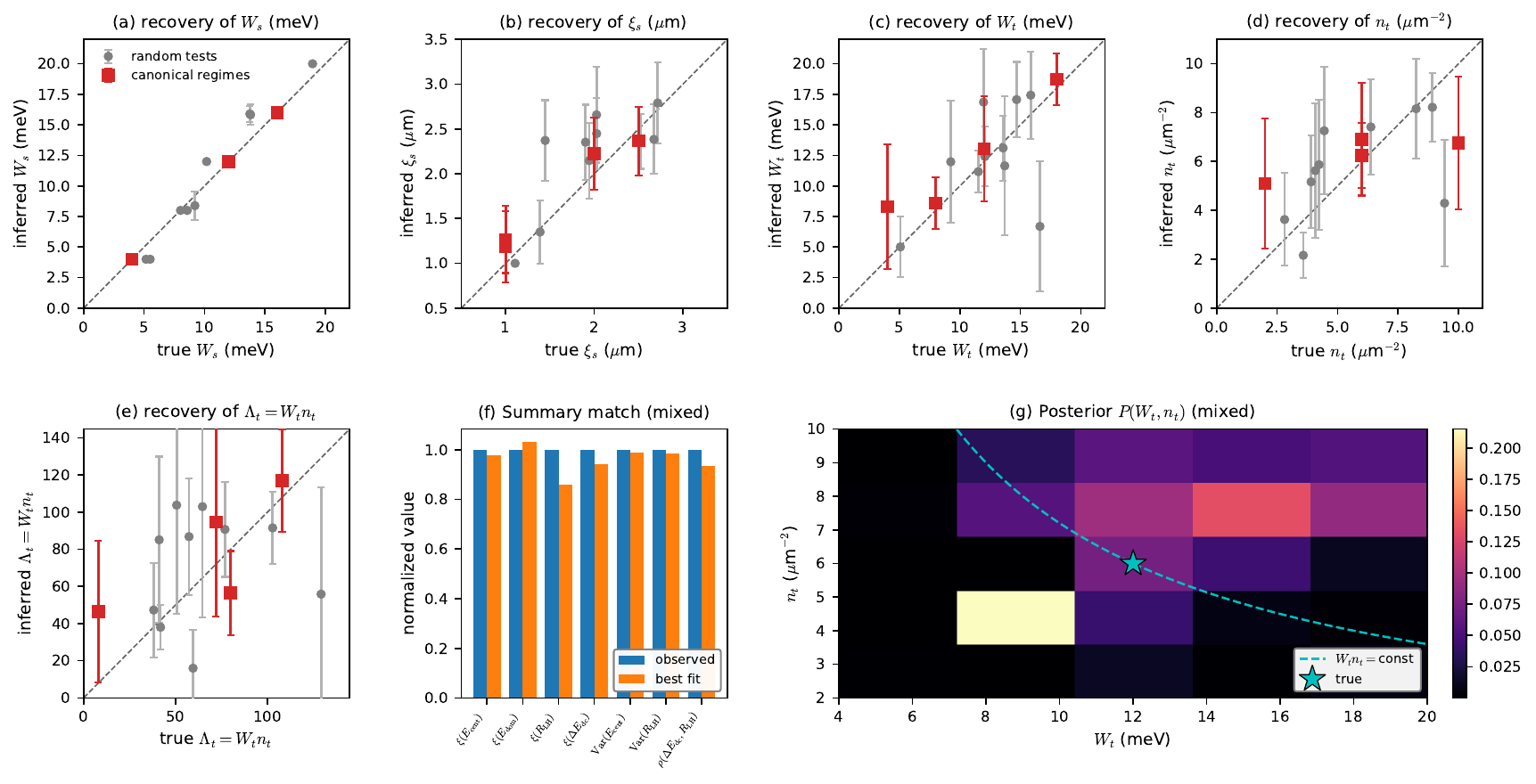}
  \caption{Inverse fitting and synthetic recovery.  (a--d) Posterior-mean
  inferred coordinate versus true value over canonical regimes (squares) and
  random test points (circles); error bars are posterior standard deviations
  and the dashed line is the identity.  (e) Recovery of the combined
  trap-activity proxy $\Lambda_t=\Wt\nt$, which clusters along the
  identity more coherently than $\Wt$ or $\nt$ individually [(c),(d)]---the
  strength--density degeneracy is an intrinsic identifiability limit, not a
  deficiency of the inverse.  (f) Observed versus best-fit summary statistics for the mixed
  regime (each component normalized by the observed magnitude).  (g) Pairwise
  posterior $P(\Wt,\nt)$ for the mixed regime, showing the trap
  strength--density degeneracy.  The dashed curve is the iso-activity line
  $\Wt\nt=$const through the true value (star), shown as a representative---not
  unique---trap-activity direction (the exponent is regime dependent;
  Appendix~\ref{app:trapdir}).}
  \label{fig:recovery}
\end{figure*}

\subsection{Summary statistics and matching}

We compress the descriptor statistics into a low-dimensional summary vector
\begin{equation}
  \begin{split}
  S = \big(&\xi_{\Ecent},\,\xi_{\Edom},\,\xi_{\RLH},\,\xi_{\dEdc},\\
  &\Var(\Ecent),\,\Var(\RLH),\,\rho(\dEdc,\RLH)\big).
  \end{split}
  \label{eq:summary}
\end{equation}
This vector is deliberately minimal and interpretable, pairing the
correlation-length hierarchy with the few covariance elements that carry
spectral-shape information.  We exclude the integrated intensity $\Iint$ from
the baseline inverse analysis.  The overall PL brightness is set largely by
experimental factors---local excitation power, collection efficiency, and
non-radiative recombination---rather than by the disorder coordinates, whereas
the spectral-shape descriptors are invariant to the intensity scale and hence
far more robust.  $\Iint$ can be added as an auxiliary descriptor once these
factors are calibrated.

The strong cross-correlation $\rho(\dEdc,\RLH)$ is a consequence of the
local-$\Edom$ split adopted for $\RLH$ in Eq.~\eqref{eq:rhl}.  Under that
convention $\RLH$ and $\dEdc$ share the same reference energy $\Edom$ and
therefore co-vary, whereas a fixed global split would decouple them and weaken
$\rho$ (Appendix~\ref{app:sensitivity}).  The inferred coordinates are thus
conditional on this descriptor convention, and comparisons across samples must
adopt the same one.  The individual
descriptor maps and their autocorrelation lengths are the primary observables.
The cross-correlation involving $\RLH$ should always be interpreted together
with the stated split convention.  As a ratio, $\RLH$ becomes noise-sensitive
when the high-energy weight (its denominator) is small.  For experimental data
its stability can be improved by using $\log\RLH$, and by masking or
regularizing pixels whose high-energy weight falls below a fixed noise floor.
This summary vector was
chosen to be minimal and interpretable rather than optimal.  Systematic feature
selection and latent-summary construction are left to future work.

For each parameter set $\Theta$ we generate synthetic cubes and compute the
same summary vector, averaging it over $n_{\mathrm{real}}$ independent disorder
realizations (here $n_{\mathrm{real}}=4$) to obtain the simulated
(forward-model) summary vector $S_{\mathrm{sim}}(\Theta)$.  The scatter of each
summary component across these realizations, $\sigma_i(\Theta)$, provides the
per-component noise model.  We floor $\sigma_i$ at the grid-median scatter to
avoid spuriously small denominators.  Increasing $n_{\mathrm{real}}$ from $4$ to
$8$ (for both the library and the observations) shifts the recovered posterior
means by a median of $0.2\sigma$, and by less than $1\sigma$ for $95\%$ of the
coordinates.  Hence $n_{\mathrm{real}}=4$ is sufficient for the present
diagnostic validation.  

Inference minimizes the
$\chi^2$ loss
\begin{equation}
  \mathcal{L}(\Theta) = \sum_i
  \frac{\left[S_i^{\mathrm{obs}}-S_i^{\mathrm{sim}}(\Theta)\right]^2}
       {\sigma_i^2},
  \label{eq:loss}
\end{equation}
where $S_i^{\mathrm{obs}}$ is the $i$th summary component computed from the single
fixed test cube being analyzed---the input whose disorder coordinates we treat as
unknown, and which in an experimental application is replaced by a measured
cube---while $S_i^{\mathrm{sim}}(\Theta)$ is the forward-model prediction at the
candidate $\Theta$, computed from cubes freshly generated for that $\Theta$ and
distinct from the test cube.  Inference thus searches for the $\Theta$ whose
simulated summary best matches the fixed observed one.

We report a posterior over a precomputed, regular $5\times5\times5\times5$
parameter grid spanning $W_s\in[4,20]$~meV, $\xi_s\in[1,3]~\mu$m,
$W_t\in[4,20]$~meV, and $n_t\in[2,10]~\mu$m$^{-2}$ (625 points)~\cite{%
tarantola2005inverse,beaumont2002approximate}.  This grid is intended for
diagnostic validation and the identifiability analysis below rather than for
high-precision parameter estimation, for which it could be refined.  Assuming a
Gaussian likelihood with a flat prior, the grid-Bayesian posterior is
\begin{equation}
  P(\Theta\,|\,S^{\mathrm{obs}}) \propto
  \exp\!\left[-\tfrac12\mathcal{L}(\Theta)\right],
  \label{eq:posterior}
\end{equation}
from which we obtain posterior means, standard deviations, and marginal and
pairwise distributions.

The loss in Eq.~\eqref{eq:loss} uses a diagonal
summary-statistics covariance, so off-diagonal correlations between summary
components are neglected.  Moreover, $\sigma_i$ captures realization scatter
(sample-to-sample plus shot noise) rather than a separately modeled
measurement-noise term, so the reported posterior widths are conditional on this
diagonal-Gaussian likelihood and the flat grid prior.  This approximation is
sufficient for the diagnostic validation performed here, but a full covariance
likelihood or a neural or approximate Bayesian computation (ABC) posterior
estimator can be incorporated in future
implementations.  This summary-statistics matching against a simulated
library is a transparent simulation-based Bayesian inference scheme, closely
related to ABC~\cite{sisson2018handbook} and to
neural simulation-based inference now being applied to spectral
fitting~\cite{barret2024simulation}.  It is sufficient for the present diagnostic
goal.

\subsection{Synthetic recovery test}

We test the inversion on the chain
$\Theta_{\mathrm{true}}\to I_{\mathrm{syn}}\to\Da\to S\to
\Theta_{\mathrm{inferred}}$, using test cubes generated from disorder
realizations that were held out of the simulation library (their random seeds
are disjoint from the library's), so the inference never sees an exact copy of
the cube it is asked to invert.
Figure~\ref{fig:recovery}(a--d) shows the recovery scatter for the four
coordinates over the four canonical regimes and ten additional random
parameter sets (sampled uniformly over the interior of the library parameter
ranges, off the library grid nodes).  The smooth-disorder coordinates $\Ws$~(a) and $\xis$~(b) are
recovered most robustly.  The trap sector is also constrained, but $\Wt$~(c) and
$\nt$~(d) are only partially identifiable on their own.  What is well constrained is
their combined trap-activity direction $\Lambda_t=\Wt\nt$ (discussed next), so
$\Wt$ and $\nt$ should be read as \emph{effective optical trap coordinates}
rather than independently determined microscopic parameters.

\subsection{Identifiability and degeneracy}

The pairwise posterior $P(\Wt,\nt)$ in Fig.~\ref{fig:recovery}(g) reveals the
origin of this uncertainty.  A small number of strong traps produces nearly the
same descriptor statistics as a larger number of weaker traps.  This is why the
posterior in Fig.~\ref{fig:recovery}(g) stretches along an iso-$\Wt\nt$ ridge,
leaving $\Wt$ and $\nt$ partially degenerate along a trap-activity direction.  We use the
simplest interpretable representative of this direction, the combined
coordinate
\begin{equation}
  \Lambda_t = \Wt\,\nt,
  \label{eq:lambda}
\end{equation}
and in the trap-rich regimes a combined trap-activity coordinate is better
constrained than either $\Wt$ or $\nt$ alone [Fig.~\ref{fig:recovery}(e)].
The precise exponent of the conserved combination is, however, regime dependent
and not sharply determined.  A dense conditional scan of $P(\Wt,\nt)$
(Appendix~\ref{app:trapdir}) yields a clean trade-off degeneracy only in the
blurred dense-trap regime, where a disorder-variance-like $\Wt^2\nt$ is in fact
marginally better conserved than $\Wt\nt$.  The trap-dominated posterior is
near-isotropic and the mixed posterior is broad.  The iso-activity curve
$\Wt\nt=$const in Fig.~\ref{fig:recovery}(g) should therefore be read as an
approximate, representative degeneracy direction rather than an exact one.

We do not assign $\Lambda_t$ a universal microscopic meaning.
It is an empirical descriptor-space trap-activity coordinate for the present
forward model, and a different combination (e.g.\ a trap variance
$\propto\Wt^2\nt$) may be more natural in other regimes.  We report this
degeneracy explicitly as an intrinsic identifiability limit of the inverse
problem rather than a deficiency of the method.  The method yields \emph{effective optical disorder
coordinates} with quantified uncertainty and explicit degeneracies, not a
unique microscopic defect reconstruction.  Figure~\ref{fig:recovery}(f)
confirms that the best-fit model reproduces the observed summary vector across
all components.

%%%%%%%%%%%%%%%%%%%%%%%%%%%%%%%%%%%%%%%%%%%%%%%%%%%%%%%%%%%%%
% (Fig.~\ref{fig:recovery} float relocated to the start of Sec.~\ref{sec:inference}
%  for better two-column placement near its first reference.)

%%%%%%%%%%%%%%%%%%%%%%%%%%%%%%%%%%%%%%%%%%%%%%%%%%%%%%%%%%%%%
\section{Disorder-coordinate diagnostics}
\label{sec:diagnostics}

\begin{figure*}[t]
  \centering
  \includegraphics[width=\textwidth]{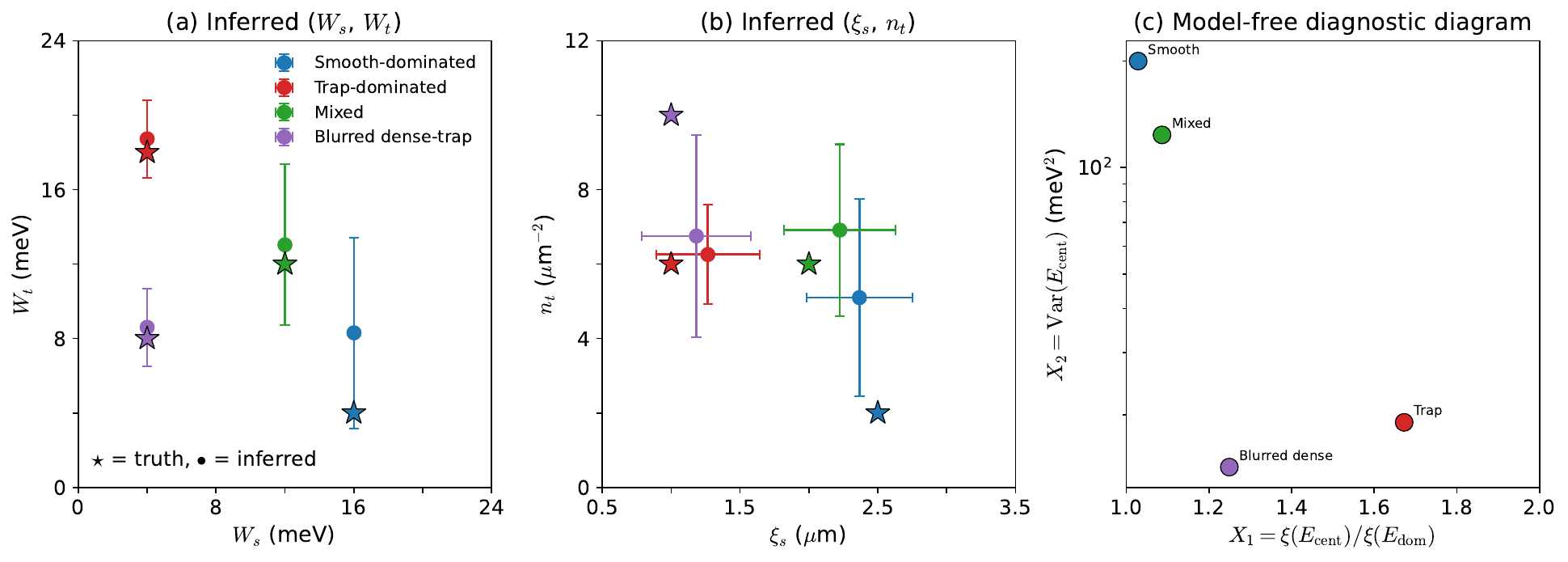}
  \caption{Disorder-coordinate diagnostics.  (a,b) Inferred effective
  coordinates of the four regimes in the $(\Ws,\Wt)$ and $(\xis,\nt)$ planes;
  circles with error bars are posterior mean $\pm$ standard deviation and
  stars are the true values.  (c) Model-free diagnostic diagram using
  $X_1=\xi(\Ecent)/\xi(\Edom)$ and $X_2=\Var(\Ecent)$ (shown on a logarithmic
  vertical scale), which separates the regimes without any inverse model.}
  \label{fig:coords}
\end{figure*}

A practical outcome of the inference is a disorder-coordinate diagnostic map.
Figure~\ref{fig:coords}(a,b)
places the four regimes in the inferred $(\Ws,\Wt)$ and $(\xis,\nt)$ planes.
The inferred coordinates (circles, with posterior-std error bars) sit close to
the true values (stars), and the regimes occupy clearly distinct regions.
Such a diagram provides a quantitative and reproducible coordinate system for
comparing samples, regions, or processing conditions.

Even without the inverse model, a purely descriptor-based diagnostic diagram
already separates the regimes.  Figure~\ref{fig:coords}(c) plots two coordinates
that are read directly from the descriptor maps, with no generative model and no
fitting to the simulation library.  The first, $X_1=\xi(\Ecent)/\xi(\Edom)$, is
the ratio of the spatial correlation lengths of the centroid and dominant-energy
maps.  The second, $X_2=\Var(\Ecent)$, is the spatial variance of the centroid
map.  Because both follow from the maps alone, we call them model-free.  The
smooth-dominated regime has $X_1\!\approx\!1$ and large $X_2$, the
trap-dominated regime has large $X_1$ and small $X_2$, and the mixed and
blurred regimes fall in between.  This model-free view is a robust first-pass
classifier that sorts the regimes but assigns neither physical coordinates nor
uncertainties.  The generative inverse then supplies those quantitative
coordinates with posterior error bars.

%%%%%%%%%%%%%%%%%%%%%%%%%%%%%%%%%%%%%%%%%%%%%%%%%%%%%%%%%%%%%

%%%%%%%%%%%%%%%%%%%%%%%%%%%%%%%%%%%%%%%%%%%%%%%%%%%%%%%%%%%%%
\section{Robustness and reusable workflow}
\label{sec:robustness}

\begin{figure*}[t]
  \centering
  \includegraphics[width=\textwidth]{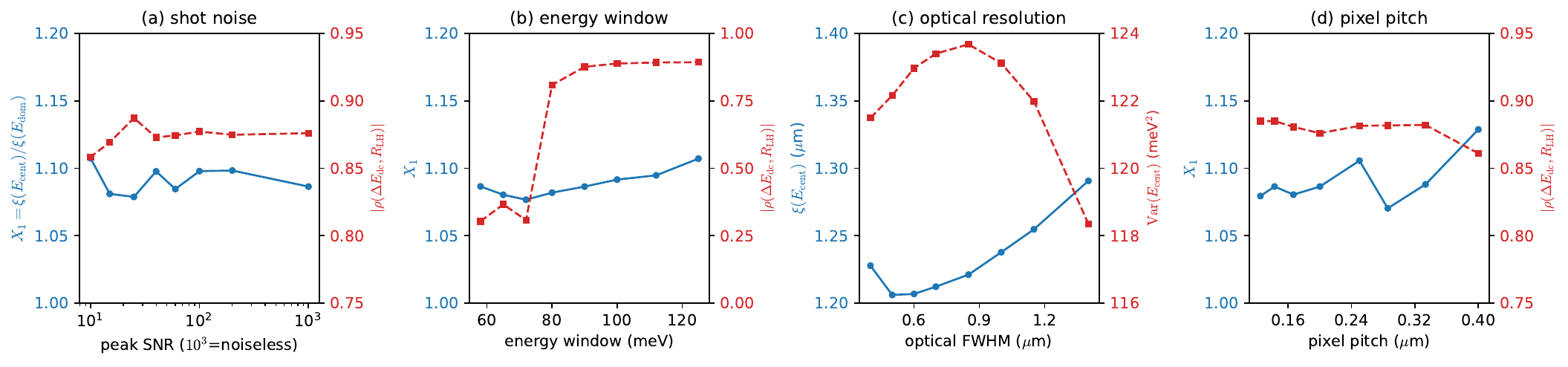}
  \caption{Robustness of the descriptor statistics for the mixed regime.  (a)
  Hierarchy ratio $X_1=\xi(\Ecent)/\xi(\Edom)$ and correlation magnitude
  $|\rho(\dEdc,\RLH)|$ versus peak shot-noise SNR.  (b) Versus energy-window
  span.  (c) $\xi(\Ecent)$ and $\Var(\Ecent)$ versus optical FWHM.  (d) $X_1$
  and $|\rho(\dEdc,\RLH)|$ versus measurement pixel pitch.}
  \label{fig:robust}
\end{figure*}

For a method to be useful as a diagnostic it must be stable against
experimental nuisances.  Figure~\ref{fig:robust} summarizes robustness tests
on the mixed regime.  The hierarchy ratio $X_1$ and the correlation magnitude
$|\rho(\dEdc,\RLH)|$ are essentially unchanged from a noiseless cube down to a
peak signal-to-noise ratio of $10$ (the shot-noise SNR
$\sqrt{N_{\mathrm{peak}}}$ of the brightest spectral channel of the brightest
pixel) [Fig.~\ref{fig:robust}(a)] and across a
factor of ${\sim}3$ in measurement pixel pitch
[Fig.~\ref{fig:robust}(d)].  Quantitatively, $X_1$ and $|\rho|$ each change by
less than ${\sim}6\%$ across the full tested SNR and pixel-pitch ranges, so the
fixed vertical scales in Fig.~\ref{fig:robust}(a,d) reflect genuine stability
rather than an axis choice.

The energy-window test
[Fig.~\ref{fig:robust}(b)] shows the expected failure mode.  A window too
narrow to contain the low-energy trap tail collapses $|\rho|$, so the window
must span the full PL band.  Once it does, the statistics are stable.  The
optical-resolution test [Fig.~\ref{fig:robust}(c)] behaves as the forward
model predicts.  Coarser resolution lengthens $\xi(\Ecent)$ and compresses the
hierarchy ratio, which is why correlation lengths are always interpreted
relative to $\lopt$.

\begin{figure}[t]
  \centering
  \includegraphics[width=0.96\columnwidth]{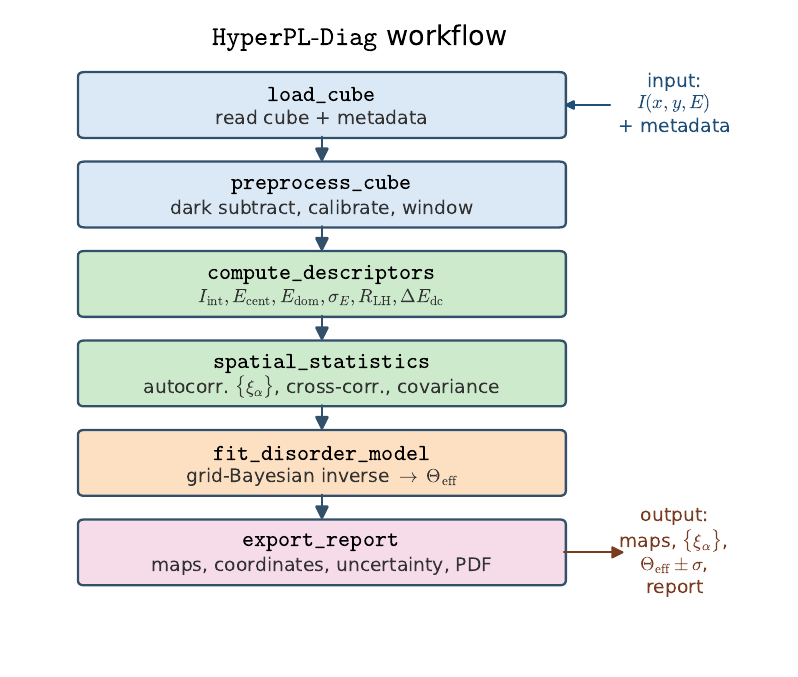}
  \caption{The \texttt{HyperPL-Diag} analysis workflow.  A hyperspectral
  cube plus metadata enters at the top; each stage is a package function, and
  the pipeline returns descriptor maps, the correlation hierarchy
  $\{\xi_\alpha\}$, the descriptor covariance, the effective disorder
  coordinates $\Thetaeff$ with uncertainty, and a summary report.}
  \label{fig:pipeline}
\end{figure}

These analyses are packaged as a reusable command-line/notebook workflow
(\texttt{HyperPL-Diag}) summarized in Fig.~\ref{fig:pipeline},
with functions spanning cube
loading, preprocessing, descriptor computation, spatial autocorrelation and
cross-correlation, correlation-length fitting, covariance estimation,
synthetic cube generation, disorder-model fitting, and report export.  The
package takes a cube plus metadata and returns descriptor maps, the
correlation hierarchy $\{\xi_\alpha\}$, the descriptor covariance, the
effective coordinates $\Thetaeff$ with uncertainty, and a summary report.

%%%%%%%%%%%%%%%%%%%%%%%%%%%%%%%%%%%%%%%%%%%%%%%%%%%%%%%%%%%%%

%%%%%%%%%%%%%%%%%%%%%%%%%%%%%%%%%%%%%%%%%%%%%%%%%%%%%%%%%%%%%

%%%%%%%%%%%%%%%%%%%%%%%%%%%%%%%%%%%%%%%%%%%%%%%%%%%%%%%%%%%%%
\section{Discussion}
\label{sec:discussion}

\paragraph{Relation to peak fitting.}
We do not argue against peak fitting, which remains valuable when the number
and identity of spectral components are well established.  For spatially
varying multi-peak moir\'e PL, however, descriptor statistics provide a more
robust and less assumption-dependent route to sample-level disorder
diagnostics, and they scale naturally to large hyperspectral datasets.  We
make this concrete in Appendix~\ref{app:peakfit}.  On a representative
synthetic spectrum, the fitted multi-Gaussian decomposition depends strongly
on the assumed number of components and on initialization, whereas the
descriptors are single deterministic numbers.

\paragraph{What the method does and does not claim.}
The descriptor fingerprint itself is model-free.  The subsequent conversion into
$\Thetaeff$ is model-assisted.  Accordingly, the inferred coordinates are
\emph{effective optical disorder coordinates},
not literal microscopic defect densities.  The forward model is minimal by
design.  The trap radius and lineshapes are fixed rather than fit, and the
$\Wt$--$\nt$ degeneracy is reported explicitly.  We do not claim that the
smooth-plus-trap model is unique.  It is the minimal calibration model used to
test identifiability and to define the effective optical coordinates.
Accordingly, the inferred posterior should be read as conditional on the chosen
descriptor convention, summary vector, and minimal forward model.  It quantifies
identifiability within this calibrated inverse problem, not a model-independent
microscopic reconstruction, and the posterior widths do not include systematic
model discrepancy.  Independent structural or
electrostatic probes---atomic force microscopy (AFM), Kelvin probe force
microscopy (KPFM), Raman, or second-harmonic generation (SHG) mapping---would
provide external validation of the optical coordinates.

\paragraph{Path to experimental data.}
The framework is built so that experimental cubes enter the same pipeline,
Eq.~\eqref{eq:pipeline}, without conceptual change.  Experimental descriptor
maps replace synthetic ones, the synthetic forward model serves as the
inference engine, and experimental samples are placed on the same
disorder-coordinate diagram.  The most informative first datasets are not the
cleanest spectra but those with a clear comparison axis---annealed versus
as-transferred, encapsulated versus bare, or clean versus bubble/strain-rich
regions.  Populating the disorder-coordinate diagram from experimental cubes to
build a disorder atlas of moir\'e PL samples is the subject of a dedicated
forthcoming study.  The present paper establishes and validates the inverse
method that such an application rests on.

\paragraph{Outlook.}
Natural extensions include gate-, power-, and polarization-resolved
descriptors, valley-resolved analysis, and machine-learning-assisted latent
descriptors~\cite{goodfellow2016deep} benchmarked against the physically
motivated set used here.  More broadly, the same descriptor-based inverse
strategy applies to hyperspectral imaging of other quantum materials,
positioning it as a general strategy for optical hidden-disorder diagnostics.

%%%%%%%%%%%%%%%%%%%%%%%%%%%%%%%%%%%%%%%%%%%%%%%%%%%%%%%%%%%%%
\section{Conclusion}
\label{sec:conclusion}

We have developed and validated a peak-decomposition-free inverse framework
that converts a hyperspectral moir\'e PL cube into a quantitative optical
metrology of effective optical disorder coordinates.  Physically motivated
descriptor maps, their spatial correlation hierarchy, and their covariance
form a disorder fingerprint.  A minimal smooth-plus-trap generative model
turns that fingerprint into effective disorder coordinates with quantified
uncertainty and explicit identifiability.  On synthetic cubes with known
ground truth the method recovers the smooth-disorder amplitude and
correlation length robustly, constrains the trap sector up to a
strength--density degeneracy captured by a combined trap-activity coordinate,
separates four canonical disorder regimes on a disorder-coordinate diagram, and
remains stable against the principal experimental nuisances.

The validated core is the foundation of a reusable analysis workflow
(\texttt{HyperPL-Diag}), and
it is structured so that experimental hyperspectral data can be incorporated
without changing the conceptual pipeline.  The central shift in perspective is
that a hyperspectral PL cube is not merely a collection of local spectra but a
spatial statistical measurement of the exciton disorder landscape.

\begin{acknowledgments}
We thank R. Kitaura, S. Urano, and N. F. Ahmad for stimulating discussions on hyperspectral PL
organization in moir\'e heterobilayers.
\end{acknowledgments}

\section*{Funding}
This work was supported by JSPS KAKENHI (Grants No.
JP25K01609 and No. JP22H05473).

\section*{Conflict of interest}
The authors declare no conflict of interest.

\section*{Ethics statement}
This research did not involve human participants, human data, human tissue, or
animals, and therefore did not require ethical approval.

\section*{Data Availability}
This paper is a methods study validated entirely on synthetic data.  No
experimental datasets were generated or analyzed.  The synthetic hyperspectral
cubes, simulation library, and inference results are reproducibly generated
with fixed random seeds by the \texttt{HyperPL-Diag} workflow and accompanying
scripts.  These scripts reproduce all figures in this paper and will be made
available upon publication.

\appendix

\section{Comparison with multi-Gaussian peak fitting}
\label{app:peakfit}

\begin{figure*}[t]
  \centering
  \includegraphics[width=\textwidth]{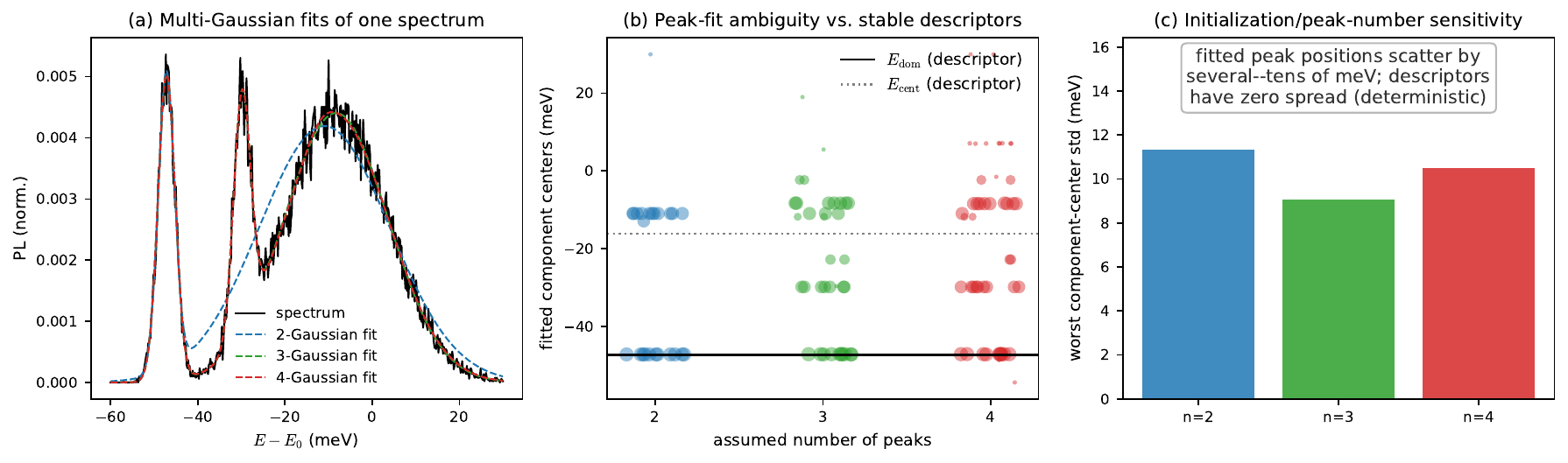}
  \caption{Descriptor robustness versus multi-Gaussian peak fitting.  (a) One
  synthetic spectrum with $n=2,3,4$ Gaussian fits.  (b) All fitted component
  centers across peak numbers and random initializations (points; marker size
  $\propto$ fitted amplitude), compared with the deterministic descriptors
  $\Edom$ (solid) and $\Ecent$ (dotted): only the dominant component is
  consistently recovered near $\Edom$, while the others scatter.  (c) Worst-case
  standard deviation of a sorted fitted-component center across initializations
  for each $n$; the descriptors have zero spread by construction.}
  \label{fig:peakfit}
\end{figure*}

To illustrate concretely why descriptor statistics are more robust than
multi-peak fitting for large, spatially varying datasets, we take one
representative synthetic spectrum (the most spectrally structured pixel of the
mixed-regime cube, with realistic shot noise) and fit it with sums of $n=2,3,4$
Gaussians from many random initializations.  Figure~\ref{fig:peakfit}(a) shows
that several peak numbers fit the spectrum comparably well, so $n$ cannot be
chosen unambiguously from the fit quality alone.  Figure~\ref{fig:peakfit}(b)
shows that the number and positions of the fitted components vary with the
assumed peak number and with the initialization.  Only the dominant (reddest)
component is consistently recovered near $\Edom$, while the remaining fitted
centers scatter over tens of meV.  Panel (c) quantifies this as the worst-case
scatter of a sorted fitted-component center across initializations.  By
contrast, the peak-decomposition-free descriptors $\Ecent$ and $\Edom$ are
single deterministic values.

This is not an argument that peak fitting is
wrong when the component structure is independently known.  It is the reason a
descriptor-based pipeline is preferable for automated, reproducible
sample-level diagnostics.  We do not claim that this example exhausts all
sophisticated peak-fitting strategies (e.g.\ information-criterion model
selection, global or regularized fitting).  Rather, it illustrates the practical
ambiguity faced by automated pixel-by-pixel decomposition when the number and
identity of components are not independently known.

\section{Sensitivity to analysis choices}
\label{app:sensitivity}

\begin{figure*}[t]
  \centering
  \includegraphics[width=\textwidth]{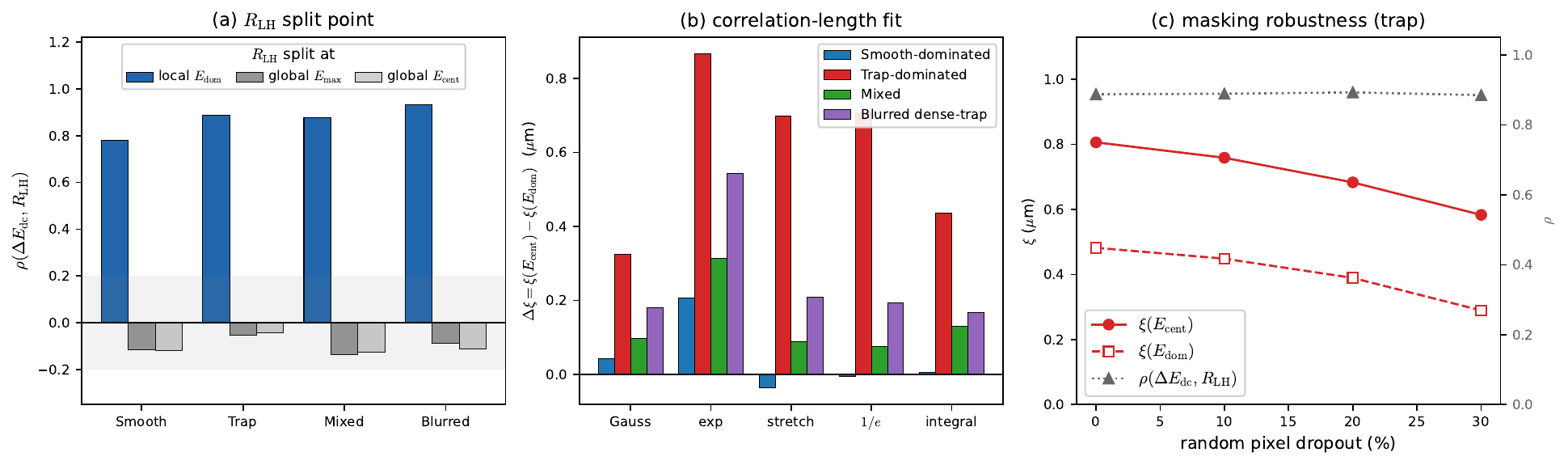}
  \caption{Sensitivity to analysis choices across the four regimes.  (a)
  $\rho(\dEdc,\RLH)$ for three $\RLH$ split points: the strong positive
  co-variation is specific to the local-$\Edom$ split (blue); a fixed global
  split (grey) gives a near-uncorrelated descriptor.  (b) The
  correlation-length gap $\Delta\xi=\xi(\Ecent)-\xi(\Edom)$ stays positive
  under all five $\xi$ definitions, largest for the trap-dominated regime.  (c)
  $\xi(\Ecent)$, $\xi(\Edom)$ and $\rho(\dEdc,\RLH)$ versus random pixel
  dropout (trap-dominated regime): the hierarchy and correlation are
  preserved.}
  \label{fig:sensitivity}
\end{figure*}

We verify that the conclusions do not hinge on three discretionary analysis
choices, using the four canonical synthetic regimes with known ground truth
(Fig.~\ref{fig:sensitivity}).

\emph{(a) $\RLH$ split point.}  Splitting $\RLH$ at the per-pixel local $\Edom$
[Eq.~\eqref{eq:rhl}] yields a strong positive $\rho(\dEdc,\RLH)$ in every
regime ($+0.78$ to $+0.93$) [Fig.~\ref{fig:sensitivity}(a)].  This co-variation is a property of the
local-$\Edom$ convention.  Here $\RLH$ and $\dEdc$ share the same dominant
energy, so both respond to the position of the dominant channel.  Splitting at
a fixed global energy instead (the map-averaged spectral maximum or centroid)
defines a distinct descriptor---the absolute spectral-weight balance about a
sample-wide reference---which is nearly uncorrelated with $\dEdc$
($|\rho|\lesssim0.14$).  The descriptor maps and their individual correlation
lengths are unchanged by the split.  Only the cross-correlation is
convention-specific, and we report it under the stated local-$\Edom$
convention.

\emph{(b) Correlation-length definition.}  Extracting $\xi$ from the
autocorrelation by a Gaussian fit, an exponential fit, a stretched exponential,
the $1/e$ crossing, or the integral correlation length all preserve the
$\xi(\Ecent)\ge\xi(\Edom)$ hierarchy, with the gap largest in the
trap-dominated regime and vanishing in the smooth-dominated regime---the same
ordering as the main text [Figs.~\ref{fig:hierarchy} and~\ref{fig:sensitivity}(b)].  The pattern of
correlation lengths, not the absolute values returned by any one estimator, is
the diagnostic.

\emph{(c) Pixel rejection.}  Rejecting low-signal pixels below a fraction of
the median integrated intensity, or randomly dropping up to $30\%$ of pixels,
leaves the hierarchy and $\rho(\dEdc,\RLH)$ essentially unchanged [Fig.~\ref{fig:sensitivity}(c)].  Absolute
correlation lengths drift only mildly as the variogram is estimated from fewer
pairs.

\section{Sensitivity to the summary-vector choice}
\label{app:summary}

\begin{figure*}[t]
  \centering
  \includegraphics[width=\textwidth]{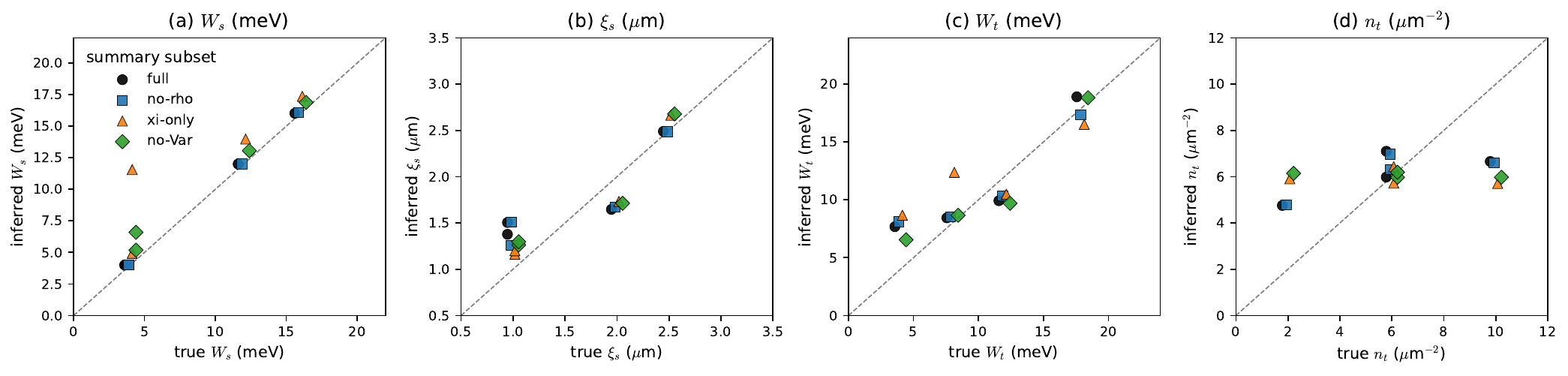}
  \caption{Sensitivity of the inferred coordinates to the summary-vector
  choice.  (a--d) Inferred-versus-true recovery for each coordinate, overlaying
  four summary subsets (full, no-$\rho$, $\xi$-only, no-Var) on the four
  canonical regimes; the dashed line is the identity.  The smooth coordinates
  are recovered consistently; the ``full'' and ``no-$\rho$'' markers nearly
  coincide everywhere, and only the hierarchy-only subset degrades, in the
  blurred dense-trap regime.}
  \label{fig:summary}
\end{figure*}

The baseline summary vector [Eq.~\eqref{eq:summary}] is one choice among many.
To verify that the inference does not hinge on it, we re-infer $\Thetaeff$ for
the four canonical regimes using subsets of the seven components, formed by
masking summary columns in the $\chi^2$ against the same library
(Fig.~\ref{fig:summary}): the full seven-component vector; dropping the
convention-defined cross-correlation $\rho(\dEdc,\RLH)$ (``no-$\rho$''); the
correlation-length hierarchy alone (``$\xi$-only''); and hierarchy plus
cross-correlation without the variances (``no-Var'').

Two results stand out.  The smooth-disorder coordinates $\Ws$ and $\xis$ are
recovered consistently across all subsets [Fig.~\ref{fig:summary}(a),(b)].  And dropping $\rho$ leaves the
inference essentially unchanged---the ``full'' and ``no-$\rho$'' points coincide
to within marker size for every regime and coordinate---so the result does not
hinge on the convention-defined cross-correlation analyzed in
Appendix~\ref{app:sensitivity}.  The correlation-length hierarchy alone already
constrains most coordinates but degrades in the blurred dense-trap regime,
where the variances carry the distinguishing information.  The covariance
components thus contribute most when traps are dense and sub-resolution.  The
trap sector retains the strength--density degeneracy under every subset [Fig.~\ref{fig:summary}(c),(d)].  The
present vector was chosen to be minimal and interpretable rather than optimal;
systematic feature selection is left to future work.

\section{Trap-sector degeneracy direction}
\label{app:trapdir}

The $\Wt$--$\nt$ degeneracy direction is not sharply defined on the five-point
inference grid, which is too coarse to fit the elongation of $P(\Wt,\nt)$.  To
probe it reliably we build a dense \emph{conditional} posterior
$P(\Wt,\nt\,|\,\Ws^\ast,\xis^\ast)$ for each trap-rich regime, fixing the smooth
coordinates at their true values and simulating the summary vector on a
$17\times17$ $(\Wt,\nt)$ grid, then matching the held-out observation with the
same diagonal-$\chi^2$ likelihood used in the main inference.  For a candidate
conserved combination $\Wt^a\nt$ we measure the unit-normalized posterior
variance of $v(a)\cdot(\log\Wt,\log\nt)$ with $v(a)=(a,1)/\lVert(a,1)\rVert$
(smaller means better conserved).  Here $a=1$ is $\Wt\nt$ and $a=2$ the
variance-like $\Wt^2\nt$.

Table~\ref{tab:trapdir} shows that the degeneracy is regime dependent.  A clean
trade-off ridge (negative posterior correlation $\rho_{\log}$ between $\log\Wt$
and $\log\nt$) appears only for the blurred dense-trap regime, where
$\Wt^2\nt$ is marginally better conserved than $\Wt\nt$.  The trap-dominated
posterior is near-isotropic (weak degeneracy) and the mixed conditional
posterior is broad and weakly positively correlated.  Thus $\Lambda_t=\Wt\nt$ is
used in the main text only as the simplest interpretable representative of trap
activity, not as a uniquely optimal coordinate.  The data do not single out a
universal exponent, and a variance-like combination is equally or more natural
where the degeneracy is sharpest.  The qualitative conclusion---that a combined
trap-activity coordinate is better constrained than $\Wt$ or $\nt$
individually---is independent of this exponent.

\begin{table}[t]
\caption{Trap-sector degeneracy direction from the dense conditional posterior
$P(\Wt,\nt\,|\,\Ws^\ast,\xis^\ast)$.  $\rho_{\log}$ is the posterior correlation
of $(\log\Wt,\log\nt)$; $\mathrm{nv}(a)$ is the unit-normalized posterior
log-variance of $\Wt^a\nt$ (smaller $=$ better conserved).  A clean trade-off
degeneracy ($\rho_{\log}<0$) comparable to the iso-activity picture appears only
for the blurred dense-trap regime, where $\Wt^2\nt$ is marginally favored.}
\label{tab:trapdir}
\begin{ruledtabular}
\begin{tabular}{lcccc}
regime & $(\Wt,\nt)$ & $\rho_{\log}$ & $\mathrm{nv}(\Wt\nt)$ &
  $\mathrm{nv}(\Wt^2\nt)$ \\
\colrule
trap-dominated     & $(18,6)$  & $-0.19$ & $0.013$ & $0.008$ \\
mixed              & $(12,6)$  & $+0.56$ & $0.365$ & $0.342$ \\
blurred dense-trap & $(8,10)$  & $-0.63$ & $0.009$ & $0.005$ \\
\end{tabular}
\end{ruledtabular}
\end{table}

\bibliographystyle{apsrev4-2}
\bibliography{refs}

@article{andrei2021marvels,
  author  = {Andrei, Eva Y. and Efetov, Dmitri K. and Jarillo-Herrero, Pablo
             and MacDonald, Allan H. and Mak, Kin Fai and Senthil, T.
             and Tutuc, Emanuel and Yazdani, Ali and Young, Andrea F.},
  title   = {{The marvels of moir\'e materials}},
  journal = {Nat. Rev. Mater.},
  volume  = {6},
  pages   = {201--206},
  year    = {2021},
  doi     = {10.1038/s41578-021-00284-1},
}

@article{mak2022semiconductor,
  author  = {Mak, Kin Fai and Shan, Jie},
  title   = {{Semiconductor moir\'e materials}},
  journal = {Nat. Nanotechnol.},
  volume  = {17},
  pages   = {686--695},
  year    = {2022},
  doi     = {10.1038/s41565-022-01165-6},
}

@article{yu2017moire,
  author  = {Yu, Hongyi and Liu, Gui-Bin and Yao, Wang},
  title   = {{Brightened spin-triplet interlayer excitons and optical selection
             rules in van der Waals heterobilayers}},
  journal = {2D Mater.},
  volume  = {5},
  pages   = {035021},
  year    = {2018},
  doi     = {10.1088/2053-1583/aac065},
}

@article{tran2019evidence,
  author  = {Tran, Kha and Moody, Galan and Wu, Fengcheng and Lu, Xiaobo
             and Choi, Junho and Kim, Kyounghwan and Rai, Amritesh
             and Sanchez, Daniel A. and Quan, Jiamin and Singh, Akshay
             and Embley, Jacob and Zepeda, André and Campbell, Marshall
             and Autry, Travis and Taniguchi, Takashi and Watanabe, Kenji
             and Lu, Nanshu and Banerjee, Sanjay K. and Silverman, Kevin L.
             and Kim, Suenne and Tutuc, Emanuel and Yang, Li
             and MacDonald, Allan H. and Li, Xiaoqin},
  title   = {{Evidence for moir\'e excitons in van der Waals heterostructures}},
  journal = {Nature},
  volume  = {567},
  pages   = {71--75},
  year    = {2019},
  doi     = {10.1038/s41586-019-0975-z},
}

@article{jin2019observation,
  author  = {Jin, Chenhao and Regan, Emma C. and Yan, Aiming
             and Utama, M. Iqbal Bakti and Wang, Danqing and Zhao, Sihan
             and Qin, Ying and Yang, Sijie and Zheng, Zhiren and Shi, Shenyang
             and Watanabe, Kenji and Taniguchi, Takashi and Tongay, Sefaattin
             and Zettl, Alex and Wang, Feng},
  title   = {{Observation of moir\'e excitons in WSe$_{2}$/WS$_{2}$
             heterostructure superlattices}},
  journal = {Nature},
  volume  = {567},
  pages   = {76--80},
  year    = {2019},
  doi     = {10.1038/s41586-019-0976-y},
}

@article{tang2020simulation,
  author  = {Tang, Yanhao and Li, Lizhong and Li, Tingxin and Xu, Yang
             and Liu, Shaowei and Barmak, Katayun and Watanabe, Kenji
             and Taniguchi, Takashi and MacDonald, Allan H. and Shan, Jie
             and Mak, Kin Fai},
  title   = {{Simulation of Hubbard model physics in WSe$_{2}$/WS$_{2}$
             moir\'e superlattices}},
  journal = {Nature},
  volume  = {579},
  pages   = {353--358},
  year    = {2020},
  doi     = {10.1038/s41586-020-2085-3},
}

@article{shimazaki2020strongly,
  author  = {Shimazaki, Yuya and Schwartz, Ido and Watanabe, Kenji
             and Taniguchi, Takashi and Kroner, Martin and Imamoglu, Atac},
  title   = {{Strongly correlated electrons and hybrid excitons in a moir\'e
             heterostructure}},
  journal = {Nature},
  volume  = {580},
  pages   = {472--477},
  year    = {2020},
  doi     = {10.1038/s41586-020-2191-2},
}

@article{seyler2019signatures,
  author  = {Seyler, Kyle L. and Rivera, Pasqual and Yu, Hongyi
             and Wilson, Nathan P. and Ray, Essance L. and Mandrus, David G.
             and Yan, Jiaqiang and Yao, Wang and Xu, Xiaodong},
  title   = {{Signatures of moir\'e-trapped valley excitons in
             MoSe$_{2}$/WSe$_{2}$ heterobilayers}},
  journal = {Nature},
  volume  = {567},
  pages   = {66--70},
  year    = {2019},
  doi     = {10.1038/s41586-019-0957-1},
}

@article{alexeev2019resonantly,
  author  = {Alexeev, Evgeny M. and Ruiz-Tijerina, David A. and Danovich, Mark
             and Hamer, Matthew J. and Terry, Daniel J. and Nayak, Pramoda K.
             and Ahn, Seongjoon and Pak, Sangyeon and Lee, Juwon
             and Sohn, Jung Inn and Molas, Maciej R. and Koperski, Maciej
             and Watanabe, Kenji and Taniguchi, Takashi and Novoselov, Kostya S.
             and Gorbachev, Roman V. and Shin, Hyeon Suk and Fal'ko, Vladimir I.
             and Tartakovskii, Alexander I.},
  title   = {{Resonantly hybridized excitons in moir\'e superlattices in
             van der Waals heterostructures}},
  journal = {Nature},
  volume  = {567},
  pages   = {81--86},
  year    = {2019},
  doi     = {10.1038/s41586-019-0986-9},
}

@article{brem2020hybridized,
  author  = {Brem, Samuel and Linder{\"a}lv, Christopher and Erhart, Paul
             and Malic, Ermin},
  title   = {{Tunable phases of moir\'e excitons in van der Waals
             heterostructures}},
  journal = {Nano Lett.},
  volume  = {20},
  pages   = {8534--8540},
  year    = {2020},
  doi     = {10.1021/acs.nanolett.0c03019},
}

@article{wu2018hubbard,
  author  = {Wu, Fengcheng and Lovorn, Timothy and Tutuc, Emanuel
             and MacDonald, Allan H.},
  title   = {{Hubbard model physics in transition metal dichalcogenide
             moir\'e bands}},
  journal = {Phys. Rev. Lett.},
  volume  = {121},
  pages   = {026402},
  year    = {2018},
  doi     = {10.1103/PhysRevLett.121.026402},
}

@article{parto2021defect,
  author  = {Parto, Kamyar and Azzam, Shaimaa I. and Banerjee, Kaustav
             and Moody, Galan},
  title   = {{Defect and strain engineering of monolayer WSe$_{2}$ enables
             site-controlled single-photon emission up to 150~K}},
  journal = {Nat. Commun.},
  volume  = {12},
  pages   = {3585},
  year    = {2021},
  doi     = {10.1038/s41467-021-23709-5},
}

@article{yu2021moire,
  author  = {Yu, Hongyi and Chen, Mingxing and Yao, Wang},
  title   = {{Giant magnetic field from moir\'e induced Berry phase in
             homobilayer semiconductors}},
  journal = {Natl. Sci. Rev.},
  volume  = {7},
  pages   = {12--20},
  year    = {2020},
  doi     = {10.1093/nsr/nwz117},
}

@article{ahmad2025hierarchical,
  author  = {Ahmad, Nurul Fariha and Urano, Yuto and Watanabe, Kenji
             and Taniguchi, Takashi and Kozawa, Daichi and Kitaura, Ryo},
  title   = {{Hierarchical spectral inhomogeneity in photoluminescence of a
             twisted MoSe$_{2}$/WSe$_{2}$ heterobilayer moir\'e superlattice
             revealed by hyperspectral mapping}},
  journal = {Appl. Phys. Lett.},
  volume  = {128},
  number  = {23},
  pages   = {231901},
  year    = {2026},
  doi     = {10.1063/5.0335306},
}

@article{splendiani2010emerging,
  author  = {Splendiani, Andrea and Sun, Liang and Zhang, Yuanbo
             and Li, Tianshu and Kim, Jonghwan and Chim, Chi-Yung
             and Galli, Giulia and Wang, Feng},
  title   = {{Emerging photoluminescence in monolayer MoS$_{2}$}},
  journal = {Nano Lett.},
  volume  = {10},
  pages   = {1271--1275},
  year    = {2010},
  doi     = {10.1021/nl903868w},
}

@article{mak2010atomically,
  author  = {Mak, Kin Fai and Lee, Changgu and Hone, James
             and Shan, Jie and Heinz, Tony F.},
  title   = {{Atomically thin MoS$_{2}$: a new direct-gap semiconductor}},
  journal = {Phys. Rev. Lett.},
  volume  = {105},
  pages   = {136805},
  year    = {2010},
  doi     = {10.1103/PhysRevLett.105.136805},
}

@article{xiao2012coupled,
  author  = {Xiao, Di and Liu, Gui-Bin and Feng, Wanxiang
             and Xu, Xiaodong and Yao, Wang},
  title   = {{Coupled spin and valley physics in monolayers of MoS$_{2}$
             and other group-{VI} dichalcogenides}},
  journal = {Phys. Rev. Lett.},
  volume  = {108},
  pages   = {196802},
  year    = {2012},
  doi     = {10.1103/PhysRevLett.108.196802},
}

@article{chernikov2014exciton,
  author  = {Chernikov, Alexey and Berkelbach, Timothy C. and Hill, Heather M.
             and Rigosi, Albert and Li, Yilei and Aslan, Ozgur Burak
             and Reichman, David R. and Hybertsen, Mark S. and Heinz, Tony F.},
  title   = {{Exciton binding energy and nonhydrogenic Rydberg series in
             monolayer WS$_{2}$}},
  journal = {Phys. Rev. Lett.},
  volume  = {113},
  pages   = {076802},
  year    = {2014},
  doi     = {10.1103/PhysRevLett.113.076802},
}

@article{berkelbachtheory,
  author  = {Berkelbach, Timothy C. and Hybertsen, Mark S.
             and Reichman, David R.},
  title   = {{Theory of neutral and charged excitons in monolayer transition
             metal dichalcogenides}},
  journal = {Phys. Rev. B},
  volume  = {88},
  pages   = {045318},
  year    = {2013},
  doi     = {10.1103/PhysRevB.88.045318},
}

@article{geim2013vdw,
  author  = {Geim, Andre K. and Grigorieva, Irina V.},
  title   = {{Van der Waals heterostructures}},
  journal = {Nature},
  volume  = {499},
  pages   = {419--425},
  year    = {2013},
  doi     = {10.1038/nature12385},
}

@article{novoselov2016heterostructures,
  author  = {Novoselov, Kostya S. and Mishchenko, Artem
             and Carvalho, Alexandra and Castro Neto, Antonio H.},
  title   = {{2D materials and van der Waals heterostructures}},
  journal = {Science},
  volume  = {353},
  pages   = {aac9439},
  year    = {2016},
  doi     = {10.1126/science.aac9439},
}

@article{mak2016photonics,
  author  = {Mak, Kin Fai and Shan, Jie},
  title   = {{Photonics and optoelectronics of 2D semiconductor transition
             metal dichalcogenides}},
  journal = {Nat. Photonics},
  volume  = {10},
  pages   = {216--226},
  year    = {2016},
  doi     = {10.1038/nphoton.2015.282},
}

@article{schaibley2016valleytronics,
  author  = {Schaibley, John R. and Yu, Hongyi and Clark, Genevieve
             and Rivera, Pasqual and Ross, Jason S. and Seyler, Kyle L.
             and Yao, Wang and Xu, Xiaodong},
  title   = {{Valleytronics in 2D materials}},
  journal = {Nat. Rev. Mater.},
  volume  = {1},
  pages   = {16055},
  year    = {2016},
  doi     = {10.1038/natrevmats.2016.55},
}

@article{wang2018colloquium,
  author  = {Wang, Gang and Chernikov, Alexey and Glazov, Mikhail M.
             and Heinz, Tony F. and Marie, Xavier and Amand, Thierry
             and Urbaszek, Bernhard},
  title   = {{Colloquium: excitons in atomically thin transition metal
             dichalcogenides}},
  journal = {Rev. Mod. Phys.},
  volume  = {90},
  pages   = {021001},
  year    = {2018},
  doi     = {10.1103/RevModPhys.90.021001},
}

@article{ciarrocchi2022excitonic,
  author  = {Ciarrocchi, Alberto and Tagarelli, Fedele and Avsar, Ahmet
             and Kis, Andras},
  title   = {{Excitonic devices with van der Waals heterostructures:
             valleytronics meets twistronics}},
  journal = {Nat. Rev. Mater.},
  volume  = {7},
  pages   = {449--464},
  year    = {2022},
  doi     = {10.1038/s41578-021-00408-7},
}

@article{rivera2015observation,
  author  = {Rivera, Pasqual and Schaibley, John R. and Jones, Aaron M.
             and Ross, Jason S. and Wu, Sanfeng and Aivazian, Genevieve
             and Klement, Philip and Seyler, Kyle and Clark, Genevieve
             and Ghimire, Nirmal J. and Yan, Jiaqiang and Mandrus, David G.
             and Yao, Wang and Xu, Xiaodong},
  title   = {{Observation of long-lived interlayer excitons in monolayer
             MoSe$_{2}$--WSe$_{2}$ heterostructures}},
  journal = {Nat. Commun.},
  volume  = {6},
  pages   = {6242},
  year    = {2015},
  doi     = {10.1038/ncomms7242},
}

@article{rivera2016valley,
  author  = {Rivera, Pasqual and Seyler, Kyle L. and Yu, Hongyi
             and Schaibley, John R. and Yan, Jiaqiang and Mandrus, David G.
             and Yao, Wang and Xu, Xiaodong},
  title   = {{Valley-polarized exciton dynamics in a 2D semiconductor
             heterostructure}},
  journal = {Science},
  volume  = {351},
  pages   = {688--691},
  year    = {2016},
  doi     = {10.1126/science.aac7820},
}

@article{nagler2017interlayer,
  author  = {Nagler, Philipp and Plechinger, Gerd and Ballottin, Mariana V.
             and Mitioglu, Anatolie and Meier, Sebastian and Paradiso, Nicola
             and Strunk, Christoph and Chernikov, Alexey
             and Christianen, Peter C. M. and Sch{\"u}ller, Christian
             and Korn, Tobias},
  title   = {{Interlayer exciton dynamics in a dichalcogenide monolayer
             heterostructure}},
  journal = {2D Mater.},
  volume  = {4},
  pages   = {025112},
  year    = {2017},
  doi     = {10.1088/2053-1583/aa7352},
}

@article{miller2017long,
  author  = {Miller, Bastian and Steinhoff, Alexander and Pano, Borja
             and Klein, Julian and Jahnke, Frank and Holleitner, Alexander
             and Wurstbauer, Ursula},
  title   = {{Long-lived direct and indirect interlayer excitons in van der
             Waals heterostructures}},
  journal = {Nano Lett.},
  volume  = {17},
  pages   = {5229--5237},
  year    = {2017},
  doi     = {10.1021/acs.nanolett.7b01304},
}

@article{wu2017theory,
  author  = {Wu, Fengcheng and Lovorn, Timothy and MacDonald, Allan H.},
  title   = {{Topological exciton bands in moir\'e heterojunctions}},
  journal = {Phys. Rev. Lett.},
  volume  = {118},
  pages   = {147401},
  year    = {2017},
  doi     = {10.1103/PhysRevLett.118.147401},
}

@article{naik2018ultraflatbands,
  author  = {Naik, Mit H. and Jain, Manish},
  title   = {{Ultraflatbands and shear solitons in moir\'e patterns of twisted
             bilayer transition metal dichalcogenides}},
  journal = {Phys. Rev. Lett.},
  volume  = {121},
  pages   = {266401},
  year    = {2018},
  doi     = {10.1103/PhysRevLett.121.266401},
}

@article{regan2020mott,
  author  = {Regan, Emma C. and Wang, Da and Jin, Chenhao
             and Utama, M. Iqbal Bakti and Gao, Beini and Wei, Xin
             and Zhao, Sihan and Zhao, Wenyu and Zhang, Zuocheng
             and Yumigeta, Kentaro and Blei, Mark and Carlstr{\"o}m, Johan D.
             and Watanabe, Kenji and Taniguchi, Takashi and Tongay, Sefaattin
             and Crommie, Michael and Zettl, Alex and Wang, Feng},
  title   = {{Mott and generalized Wigner crystal states in
             WSe$_{2}$/WS$_{2}$ moir\'e superlattices}},
  journal = {Nature},
  volume  = {579},
  pages   = {359--363},
  year    = {2020},
  doi     = {10.1038/s41586-020-2092-4},
}

@article{weston2020atomic,
  author  = {Weston, Astrid and Zou, Yichao and Enaldiev, Vladimir
             and Summerfield, Alex and Clark, Nicholas and Z{\'o}lyomi, Viktor
             and Graham, Abigail and Yelgel, Celal and Magorrian, Samuel
             and Zhou, Mingwei and Zultak, Johanna and Hopkinson, David
             and Barinov, Alexei and Bointon, Thomas H. and Kretinin, Andrey
             and Wilson, Neil R. and Beton, Peter H. and Fal'ko, Vladimir I.
             and Haigh, Sarah J. and Gorbachev, Roman},
  title   = {{Atomic reconstruction in twisted bilayers of transition metal
             dichalcogenides}},
  journal = {Nat. Nanotechnol.},
  volume  = {15},
  pages   = {592--597},
  year    = {2020},
  doi     = {10.1038/s41565-020-0682-9},
}

@article{mcgilly2020visualization,
  author  = {McGilly, Leo J. and Kerelsky, Alexander and Finney, Nathan R.
             and Shapovalov, Konstantin and Shih, En-Min and Ghiotto, Augusto
             and Zeng, Yihang and Moore, Samuel L. and Wu, Wenjing
             and Bai, Yusong and Watanabe, Kenji and Taniguchi, Takashi
             and Stengel, Massimiliano and Zhou, Lin and Hone, James
             and Zhu, Xiaoyang and Basov, Dmitri N. and Dean, Cory
             and Dreyer, Cyrus E. and Pasupathy, Abhay N.},
  title   = {{Visualization of moir\'e superlattices}},
  journal = {Nat. Nanotechnol.},
  volume  = {15},
  pages   = {580--584},
  year    = {2020},
  doi     = {10.1038/s41565-020-0708-3},
}

@article{shabani2021deep,
  author  = {Shabani, Sara and Halbertal, Dorri and Wu, Wenjing
             and Chen, Mingxing and Liu, Song and Hone, James
             and Yao, Wang and Basov, D. N.
             and Zhu, Xiaoyang and Pasupathy, Abhay N.},
  title   = {{Deep moir\'e potentials in twisted transition metal dichalcogenide
             bilayers}},
  journal = {Nat. Phys.},
  volume  = {17},
  pages   = {720--725},
  year    = {2021},
  doi     = {10.1038/s41567-021-01174-7},
}

@article{conley2013bandgap,
  author  = {Conley, Hiram J. and Wang, Bin and Ziegler, Jed I.
             and Haglund, Richard F. and Pantelides, Sokrates T.
             and Bolotin, Kirill I.},
  title   = {{Bandgap engineering of strained monolayer and bilayer MoS$_{2}$}},
  journal = {Nano Lett.},
  volume  = {13},
  pages   = {3626--3630},
  year    = {2013},
  doi     = {10.1021/nl4014748},
}

@article{he2015single,
  author  = {He, Yu-Ming and Clark, Genevieve and Schaibley, John R.
             and He, Yu and Chen, Ming-Cheng and Wei, Yu-Jia and Ding, Xing
             and Zhang, Qiang and Yao, Wang and Xu, Xiaodong and Lu, Chao-Yang
             and Pan, Jian-Wei},
  title   = {{Single quantum emitters in monolayer semiconductors}},
  journal = {Nat. Nanotechnol.},
  volume  = {10},
  pages   = {497--502},
  year    = {2015},
  doi     = {10.1038/nnano.2015.75},
}

@article{koperski2015single,
  author  = {Koperski, Maciej and Nogajewski, Karol and Arora, Ashish
             and Cherkez, Vitalii and Mallet, Pierre and Veuillen, Jean-Yves
             and Marcus, Jason and Kossacki, Piotr and Potemski, Marek},
  title   = {{Single photon emitters in exfoliated WSe$_{2}$ structures}},
  journal = {Nat. Nanotechnol.},
  volume  = {10},
  pages   = {503--506},
  year    = {2015},
  doi     = {10.1038/nnano.2015.67},
}

@article{srivastava2015optically,
  author  = {Srivastava, Ajit and Sidler, Meinrad and Allain, Adrien V.
             and Lembke, Dominik S. and Kis, Andras and Imamoglu, Atac},
  title   = {{Optically active quantum dots in monolayer WSe$_{2}$}},
  journal = {Nat. Nanotechnol.},
  volume  = {10},
  pages   = {491--496},
  year    = {2015},
  doi     = {10.1038/nnano.2015.60},
}

@article{tonndorf2015single,
  author  = {Tonndorf, Philipp and Schmidt, Robert and Schneider, Robert
             and Kern, Johannes and Buscema, Michele and Steele, Gary A.
             and Castellanos-Gomez, Andres and van der Zant, Herre S. J.
             and Michaelis de Vasconcellos, Steffen and Bratschitsch, Rudolf},
  title   = {{Single-photon emission from localized excitons in an atomically
             thin semiconductor}},
  journal = {Optica},
  volume  = {2},
  pages   = {347--352},
  year    = {2015},
  doi     = {10.1364/OPTICA.2.000347},
}

@article{branny2017deterministic,
  author  = {Branny, Artur and Kumar, Santosh and Proux, Raphael
             and Gerardot, Brian D.},
  title   = {{Deterministic strain-induced arrays of quantum emitters in a
             two-dimensional semiconductor}},
  journal = {Nat. Commun.},
  volume  = {8},
  pages   = {15053},
  year    = {2017},
  doi     = {10.1038/ncomms15053},
}

@article{kulig2018exciton,
  author  = {Kulig, Marvin and Zipfel, Jonas and Nagler, Philipp
             and Blanter, Sofia and Sch{\"u}ller, Christian and Korn, Tobias
             and Paradiso, Nicola and Glazov, Mikhail M. and Chernikov, Alexey},
  title   = {{Exciton diffusion and halo effects in monolayer semiconductors}},
  journal = {Phys. Rev. Lett.},
  volume  = {120},
  pages   = {207401},
  year    = {2018},
  doi     = {10.1103/PhysRevLett.120.207401},
}

@article{zipfel2020exciton,
  author  = {Zipfel, Jonas and Kulig, Marvin and Perea-Caus{\'i}n, Raul
             and Brem, Samuel and Ziegler, Jonas D. and Rosati, Roberto
             and Taniguchi, Takashi and Watanabe, Kenji and Glazov, Mikhail M.
             and Malic, Ermin and Chernikov, Alexey},
  title   = {{Exciton diffusion in monolayer semiconductors with suppressed
             disorder}},
  journal = {Phys. Rev. B},
  volume  = {101},
  pages   = {115430},
  year    = {2020},
  doi     = {10.1103/PhysRevB.101.115430},
}

@article{moody2015intrinsic,
  author  = {Moody, Galan and Dass, Chandriker Kavir and Hao, Kai
             and Chen, Chang-Hsiao and Li, Lain-Jong and Singh, Akshay
             and Tran, Kha and Clark, Genevieve and Xu, Xiaodong
             and Bergh{\"a}user, Gunnar and Malic, Ermin and Knorr, Andreas
             and Li, Xiaoqin},
  title   = {{Intrinsic homogeneous linewidth and broadening mechanisms of
             excitons in monolayer transition metal dichalcogenides}},
  journal = {Nat. Commun.},
  volume  = {6},
  pages   = {8315},
  year    = {2015},
  doi     = {10.1038/ncomms9315},
}

@article{cadiz2017excitonic,
  author  = {Cadiz, Fabian and Courtade, Emmanuel and Robert, Cedric
             and Wang, Gang and Shen, Ying and Cai, Han and Taniguchi, Takashi
             and Watanabe, Kenji and Carrere, Helene and Lagarde, Delphine
             and Manca, Maxime and Amand, Thierry and Renucci, Pierre
             and Tongay, Sefaattin and Marie, Xavier and Urbaszek, Bernhard},
  title   = {{Excitonic linewidth approaching the homogeneous limit in
             MoS$_{2}$-based van der Waals heterostructures}},
  journal = {Phys. Rev. X},
  volume  = {7},
  pages   = {021026},
  year    = {2017},
  doi     = {10.1103/PhysRevX.7.021026},
}

@article{anderson1958absence,
  author  = {Anderson, Philip W.},
  title   = {{Absence of diffusion in certain random lattices}},
  journal = {Phys. Rev.},
  volume  = {109},
  pages   = {1492--1505},
  year    = {1958},
  doi     = {10.1103/PhysRev.109.1492},
}

@article{abrahams1979scaling,
  author  = {Abrahams, Elihu and Anderson, Philip W.
             and Licciardello, Donald C. and Ramakrishnan, Thirupataiah V.},
  title   = {{Scaling theory of localization: absence of quantum diffusion in
             two dimensions}},
  journal = {Phys. Rev. Lett.},
  volume  = {42},
  pages   = {673--676},
  year    = {1979},
  doi     = {10.1103/PhysRevLett.42.673},
}

@article{lee1985disordered,
  author  = {Lee, Patrick A. and Ramakrishnan, Thirupataiah V.},
  title   = {{Disordered electronic systems}},
  journal = {Rev. Mod. Phys.},
  volume  = {57},
  pages   = {287--337},
  year    = {1985},
  doi     = {10.1103/RevModPhys.57.287},
}

@article{kramer1993localization,
  author  = {Kramer, Bernhard and MacKinnon, Angus},
  title   = {{Localization: theory and experiment}},
  journal = {Rep. Prog. Phys.},
  volume  = {56},
  pages   = {1469--1564},
  year    = {1993},
  doi     = {10.1088/0034-4885/56/12/001},
}

@article{kennes2021moire,
  author  = {Kennes, Dante M. and Claassen, Martin and Xian, Lede
             and Georges, Antoine and Millis, Andrew J. and Hone, James
             and Dean, Cory R. and Basov, D. N. and Pasupathy, Abhay N.
             and Rubio, Angel},
  title   = {{Moir\'e heterostructures as a condensed-matter quantum
             simulator}},
  journal = {Nat. Phys.},
  volume  = {17},
  pages   = {155--163},
  year    = {2021},
  doi     = {10.1038/s41567-020-01154-3},
}

@article{bai2020excitons,
  author  = {Bai, Yusong and Zhou, Lin and Wang, Jue and Wu, Wenjing
             and McGilly, Leo J. and Halbertal, Dorri and Lo, Chiu Fan Bowen
             and Liu, Fang and Ardelean, Jenny and Rivera, Pasqual
             and Finney, Nathan R. and Yang, Xu-Chen and Basov, Dmitri N.
             and Yao, Wang and Xu, Xiaodong and Hone, James
             and Pasupathy, Abhay N. and Zhu, X.-Y.},
  title   = {{Excitons in strain-induced one-dimensional moir\'e potentials at
             transition metal dichalcogenide heterojunctions}},
  journal = {Nat. Mater.},
  volume  = {19},
  pages   = {1068--1073},
  year    = {2020},
  doi     = {10.1038/s41563-020-0730-8},
}

@article{komsa2015native,
  author  = {Komsa, Hannu-Pekka and Krasheninnikov, Arkady V.},
  title   = {{Native defects in bulk and monolayer MoS$_{2}$ from first
             principles}},
  journal = {Phys. Rev. B},
  volume  = {91},
  pages   = {125304},
  year    = {2015},
  doi     = {10.1103/PhysRevB.91.125304},
}

@article{lindlau2018role,
  author  = {Lindlau, Jessica and Selig, Malte and Neumann, Andre
             and Colombier, L{\'e}o and F{\"o}rste, Jonathan and Funk, Victor
             and F{\"o}rg, Michael and Kim, Jonghwan and Bergh{\"a}user, Gunnar
             and Taniguchi, Takashi and Watanabe, Kenji and Wang, Feng
             and Malic, Ermin and H{\"o}gele, Alexander},
  title   = {{The role of momentum-dark excitons in the elementary optical
             response of bilayer WSe$_{2}$}},
  journal = {Nat. Commun.},
  volume  = {9},
  pages   = {2586},
  year    = {2018},
  doi     = {10.1038/s41467-018-04877-3},
}

@misc{wakabayashi2025descriptor,
  author        = {Wakabayashi, Katsunori},
  title         = {{Descriptor Covariance and Correlation Hierarchy in Moir\'e
                   Exciton Photoluminescence}},
  year          = {2026},
  eprint        = {2606.06780},
  archivePrefix = {arXiv},
}

@article{beaumont2002approximate,
  title = {Approximate Bayesian computation in population genetics},
  author = {Beaumont, Mark A and Zhang, Wenyang and Balding, David J},
  journal = {Genetics},
  volume = {162},
  number = {4},
  pages = {2025--2035},
  year = {2002},
  publisher = {Oxford University Press}
}

@incollection{sisson2018handbook,
  title = {Overview of {ABC}},
  author = {Sisson, Scott A. and Fan, Yanan and Beaumont, Mark A.},
  booktitle = {Handbook of Approximate Bayesian Computation},
  pages = {3--54},
  year = {2018},
  publisher = {Chapman and Hall/CRC}
}

@book{cressie1993statistics,
  title = {Statistics for Spatial Data},
  author = {Cressie, Noel A. C.},
  year = {1993},
  publisher = {Wiley},
  address = {New York}
}

@book{tarantola2005inverse,
  title = {Inverse Problem Theory and Methods for Model Parameter Estimation},
  author = {Tarantola, Albert},
  year = {2005},
  publisher = {Society for Industrial and Applied Mathematics},
  address = {Philadelphia}
}

@book{goodfellow2016deep,
  title = {Deep Learning},
  author = {Goodfellow, Ian and Bengio, Yoshua and Courville, Aaron},
  year = {2016},
  publisher = {MIT Press},
  address = {Cambridge, MA}
}

@article{raja2019dielectric,
  title = {Dielectric disorder in two-dimensional materials},
  author = {Raja, Archana and Waldecker, Lutz and Zipfel, Jonas and Cho, Yeongsu
           and Brem, Samuel and Ziegler, Jonas D. and Kulig, Marvin
           and Taniguchi, Takashi and Watanabe, Kenji and Malic, Ermin
           and Heinz, Tony F. and Berkelbach, Timothy C. and Chernikov, Alexey},
  journal = {Nat. Nanotechnol.},
  volume = {14},
  number = {9},
  pages = {832--837},
  year = {2019},
  doi = {10.1038/s41565-019-0520-0},
}

@article{regan2022emerging,
  author  = {Regan, Emma C. and Wang, Danqing and Paik, Eunice Y. and Zeng,
             Yongxin and Zhang, Long and Zhu, Jihang and MacDonald, Allan H.
             and Deng, Hui and Wang, Feng},
  title   = {{Emerging exciton physics in transition metal dichalcogenide
             heterobilayers}},
  journal = {Nat. Rev. Mater.},
  volume  = {7},
  pages   = {778--795},
  year    = {2022},
  doi     = {10.1038/s41578-022-00440-1},
}

@article{blundo2024localisation,
  author  = {Blundo, Elena and Tuzi, Federico and Cianci, Salvatore and Cuccu,
             Marzia and Olkowska-Pucko, Katarzyna and Kipczak, {\L}ucja and
             Contestabile, Giorgio and Miriametro, Antonio and Felici, Marco and
             Pettinari, Giorgio and Taniguchi, Takashi and Watanabe, Kenji and
             Babi{\'n}ski, Adam and Molas, Maciej R. and Polimeni, Antonio},
  title   = {{Localisation-to-delocalisation transition of moir\'e excitons in
             WSe$_2$/MoSe$_2$ heterostructures}},
  journal = {Nat. Commun.},
  volume  = {15},
  pages   = {1057},
  year    = {2024},
  doi     = {10.1038/s41467-024-44739-9},
}

@article{fang2023localization,
  author  = {Fang, Hanlin and Lin, Qiaoling and Zhang, Yi and Thompson, Joshua
             and Xiao, Sanshui and Sun, Zhipei and Malic, Ermin and Dash,
             Saroj P. and Wieczorek, Witlef},
  title   = {{Localization and interaction of interlayer excitons in
             MoSe$_2$/WSe$_2$ heterobilayers}},
  journal = {Nat. Commun.},
  volume  = {14},
  pages   = {6910},
  year    = {2023},
  doi     = {10.1038/s41467-023-42710-8},
}

@article{brotonsgisbert2024interlayer,
  author  = {Brotons-Gisbert, Mauro and Gerardot, Brian D. and Holleitner,
             Alexander W. and Wurstbauer, Ursula},
  title   = {{Interlayer and moir\'e excitons in atomically thin double
             layers: from individual quantum emitters to degenerate
             ensembles}},
  journal = {MRS Bull.},
  volume  = {49},
  number  = {9},
  pages   = {914--931},
  year    = {2024},
  doi     = {10.1557/s43577-024-00772-z},
}

@article{barret2024simulation,
  author  = {Barret, Didier and Dupourqu\'e, Simon},
  title   = {{Simulation-Based Inference with Neural Posterior Estimation
             applied to X-ray spectral fitting: Demonstration of working
             principles down to the Poisson regime}},
  journal = {Astron. Astrophys.},
  volume  = {686},
  pages   = {A133},
  year    = {2024},
  doi     = {10.1051/0004-6361/202449214},
}

@article{shinokita2022valley,
  author  = {Shinokita, Keisuke and Watanabe, Kenji and Taniguchi, Takashi and
             Matsuda, Kazunari},
  title   = {{Valley Relaxation of the Moir\'e Excitons in a WSe$_2$/MoSe$_2$
             Heterobilayer}},
  journal = {ACS Nano},
  volume  = {16},
  number  = {10},
  pages   = {16862--16868},
  year    = {2022},
  doi     = {10.1021/acsnano.2c06813},
}

@misc{alfrey2026revealing,
  author        = {Alfrey, Adam and Tait, Cole and Taniguchi, Takashi and
                   Watanabe, Kenji and Cundiff, Steven T.},
  title         = {{Revealing Strain and Disorder in Transition-Metal
                   Dichalcogenides Using Hyperspectral Photoluminescence
                   Imaging}},
  year          = {2026},
  eprint        = {2604.01412},
  archivePrefix = {arXiv},
}

@article{eda2011photoluminescence,
  author  = {Eda, Goki and Yamaguchi, Hisato and Voiry, Damien and Fujita,
             Takeshi and Chen, Mingwei and Chhowalla, Manish},
  title   = {{Photoluminescence from Chemically Exfoliated MoS$_2$}},
  journal = {Nano Lett.},
  volume  = {11},
  number  = {12},
  pages   = {5111--5116},
  year    = {2011},
  doi     = {10.1021/nl201874w},
}

@article{okada2018direct,
  author  = {Okada, Mitsuhiro and Kutana, Alex and Kureishi, Yusuke and
             Kobayashi, Yu and Saito, Yuika and Saito, Tetsuki and Watanabe,
             Kenji and Taniguchi, Takashi and Gupta, Sunny and Miyata, Yasumitsu
             and Yakobson, Boris I. and Shinohara, Hisanori and Kitaura, Ryo},
  title   = {{Direct and Indirect Interlayer Excitons in a van der Waals
             Heterostructure of hBN/WS$_2$/MoS$_2$/hBN}},
  journal = {ACS Nano},
  volume  = {12},
  number  = {3},
  pages   = {2498--2505},
  year    = {2018},
  doi     = {10.1021/acsnano.7b08253},
}

@article{schaibley2022localized,
  author  = {Mahdikhanysarvejahany, Fateme and Shanks, Daniel N. and Klein,
             Mathew and Wang, Qian and Koehler, Michael R. and Mandrus, David G.
             and Taniguchi, Takashi and Watanabe, Kenji and Monti, Oliver L. A.
             and LeRoy, Brian J. and Schaibley, John R.},
  title   = {{Localized interlayer excitons in MoSe$_2$--WSe$_2$ heterostructures
             without a moir\'e potential}},
  journal = {Nat. Commun.},
  volume  = {13},
  pages   = {5354},
  year    = {2022},
  doi     = {10.1038/s41467-022-33082-6},
}

\end{document}